%% file: main.tex
\documentclass[letterpaper,twocolumn,10pt]{article}
\usepackage{usenix-2020-09}
\usepackage{amsmath}
\usepackage{diagbox}
\usepackage[english]{babel}
\usepackage{color,colortbl,multirow,hyphenat}
\usepackage{tikz,hyperref,xspace,booktabs}
\usepackage{tabularx,enumitem}
\usepackage[labelformat=simple, textfont=normalfont]{subcaption}
\usepackage{multirow}
\usepackage[normalem]{ulem}
\usepackage{amsfonts,algorithm}
\usepackage{algorithm}
\usepackage{algpseudocode}
\usepackage{titlesec}
\usepackage{appendix}
\usepackage[capitalize]{cleveref}
\usepackage{booktabs}
\usepackage{gensymb}
\usepackage{array}
\usepackage{booktabs}
\usepackage{graphicx}
\usepackage{threeparttable}
\usepackage{wasysym}

\newcommand{\ie}{{\itshape i.e.}\xspace}

\newcommand{\parabreak}{\vspace*{2.00ex minus 0.25ex}\noindent}

\newcommand{\parahead}[1]{\vspace*{1.25ex plus 0.25ex minus 0.25ex}\noindent{}{\bfseries #1.}}

\definecolor{Gray}{gray}{0.9}
\definecolor{LightGreen}{rgb}{0.88,1,0.88}
\definecolor{LightYellow}{rgb}{1,1,0.8}
\definecolor{LightOrange}{rgb}{1,0.85,0.8}
\definecolor{LightRed}{rgb}{1,0.80,0.80}

\newcommand{\systemname}{WaveFlex}
\newcommand{\systemnames}{WaveFlex's}

\DeclareCaptionLabelSeparator{endash}{--- }
\newcolumntype{P}[1]{>{\centering\arraybackslash}p{#1}}
\newcolumntype{M}[1]{>{\centering\arraybackslash}m{#1}}

\DeclareMathOperator*{\argmax}{argmax} 

\begin{document}
\date{}

\title{\Large \bf WaveFlex: A Smart Surface for Private CBRS Wireless Cellular Networks}



\author{
    \begin{tabular}{c}
        {\rm Fan Yi\textsuperscript{1}}, 
        {\rm Kun Woo Cho\textsuperscript{1}}, 
        {\rm Yaxiong Xie\textsuperscript{2}}, 
        {\rm Kyle Jamieson\textsuperscript{1}} \\
        \\
        \textsuperscript{1}Princeton University,
        \textsuperscript{2}University at Buffalo
    \end{tabular}
}

\maketitle

\begin{abstract}
\input{text/0-abstract}
\end{abstract}

\pagestyle{plain}
\thispagestyle{empty}

\input{text/1-intro.tex}
\input{text/2-primer}

\input{text/3-related.tex}
\input{text/5-design.tex}

\input{text/6-impl.tex}
\input{text/7-eval.tex}

\input{text/9-concl.tex}
\input{text/10-acks.tex}

\clearpage

\pagenumbering{roman}
\let\oldbibliography\thebibliography
\renewcommand{\thebibliography}[1]{%
  \oldbibliography{#1}%
  \setlength{\parskip}{0pt}%
  \setlength{\itemsep}{0pt}%
}
\begin{raggedright}
\bibliographystyle{concise2}
\bibliography{references}
\end{raggedright}


\appendix
\input{text/appendix_file.tex}

\end{document}

%% file: text/0-abstract.tex
We present the design and implementation of 
\textbf{\systemname{}}, the first
smart surface that enhances Private LTE/5G networks operating 
under the shared-license framework in the
Citizens Broadband Radio Service frequency band. 
WaveFlex works in the presence of
frequency diversity: multiple nearby base stations 
operating on different frequencies, as dictated by a Spectrum 
Access System coordinator.  It also handles time dynamism:
due to the dynamic sharing rules of the band, base stations 
occasionally switch channels, especially 
when priority users enter the network. Finally, 
WaveFlex operates independently of the network itself,
not requiring access to nor modification of the 
base station or mobile users, yet it
remain compliant with and effective on prevailing cellular 
protocols.  We have designed and fabricated WaveFlex on a custom
multi-layer PCB, software defined radio-based network monitor,
and supporting control software and hardware.  Our experimental
evaluation benchmarks an operational Private LTE network running
at full line rate.  Results demonstrate an 8.50~dB average SNR gain, 
and an average throughput gain of 4.36 Mbps
for a single small cell, and 3.19 Mbps for four small cells,
in a realistic indoor office scenario.

%% file: text/1-intro.tex
\section{Introduction}

Starting in 1986 with the original design of the predecessor of
WaveLAN, and continuing to 802.11 Wi-Fi, wireless networks deployed 
by enterprises \cite{denseAP, dirc-sigcomm09}
and home users have operated on a technology trajectory determined by 
Wi-Fi hardware, lagging years behind its wireless cellular network cousin.

That status quo is poised to change, however, in the context of 
the \emph{``Industry 4.0''} manufacturing, retail, and general 
business trend, and specifically with the advent of
locally\hyp{}managed enterprise 5G deployments
\cite{Peterson5G}.
These \emph{Private LTE\fshyp{}5G} cellular 
networks have recently emerged as a serious competitor that may
soon supplant traditional Wi-Fi, due to their superior capabilities.
Networks that scale in density to millions of nodes per square mile
will support massive Internet of Things deployment of embedded
devices and tags.  
Networks that scale
latency down to milliseconds will support real\hyp{}time control applications.
And networks that scale data rates and capacity to 
multi\hyp{}Gbit\fshyp{}second
and Tbit\fshyp{}second aggregate throughput per square mile will 
fully connect users. 

Private cellular networks can operate in a number of bands in the radio
spectrum, each of which may be accessed under a
\emph{licensed}, \emph{unlicensed}, or \emph{shared} basis,
according to the relevant
government regulations.  Shared spectrum access---in the 
\emph{Citizens Broadband Radio Service (CBRS)} band in the 
U.S. \cite{fcc35gHz}---is making grassroots, bottom\hyp{}up deployment
of Private 5G networks possible, spurring innovation in a way 
similar to Wi-Fi in the early 2000s.
CBRS offers a unique opportunity for enterprises to deploy Private 
LTE and 5G networks without needing to acquire expensive, 
exclusive-use licensed spectrum, 
even further catalyzing innovation.

\begin{figure}[hb]
\begin{subfigure}[b]{.53\linewidth}
\centering
\includegraphics[width=1\linewidth]{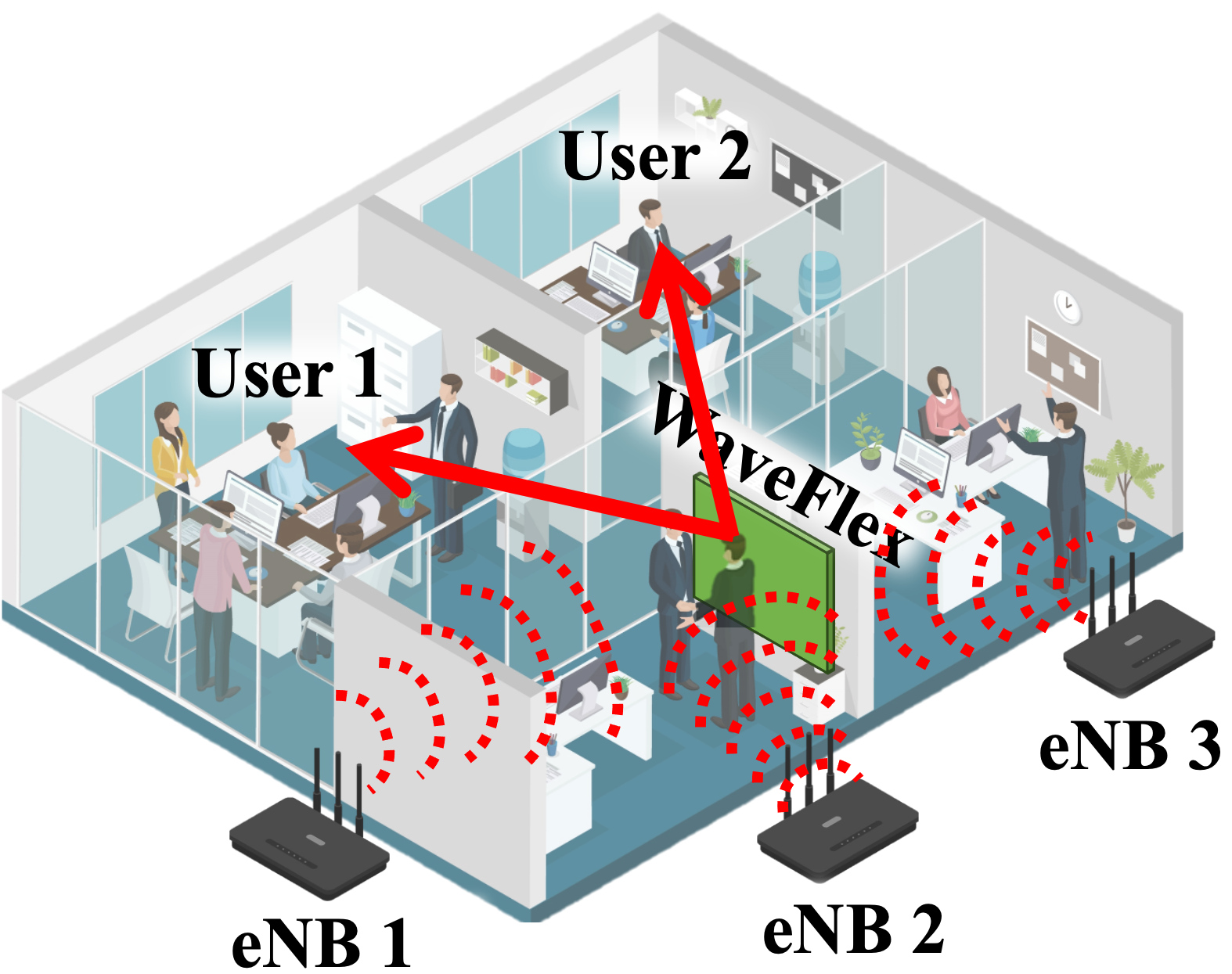}
\caption{Optimizes multi\hyp{}eNB\fshyp{}multi\hyp{}user Private 5G networks.}
\label{f:usecase2}
\end{subfigure}
\hfill
\begin{subfigure}[b]{.46\linewidth}
\centering
\includegraphics[width=1\linewidth]{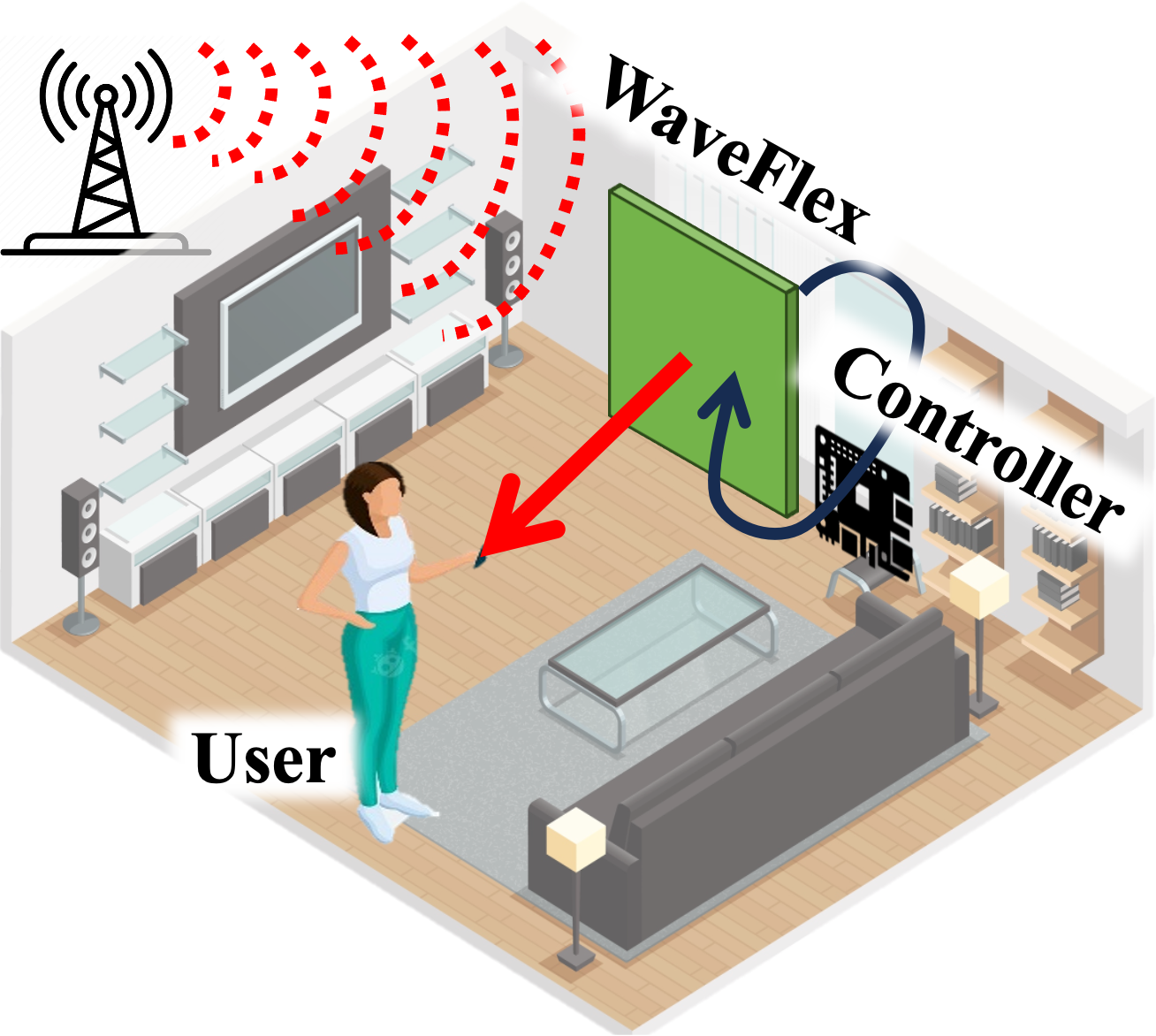}
\caption{Works independently of the network.}
\label{f:usecase3}
\end{subfigure}
\caption{\textbf{\systemname{}:} A surface for CBRS Private LTE/5G.}
\label{f:usecase}
\end{figure}

\parabreak{}This paper describes the design and implementation of
\textit{\textbf{\systemname{}}}, the first practical smart surface that 
works at full network line rate to enhance the performance of 
Private LTE/5G\footnote{\cref{s:concl} comapres 
Private LTE with Private 5G, and outlines \systemnames{} 
capabilities and limitations relative to each.}
wireless cellular networks 
operating in the CBRS band, and sharing that band with 
incumbent priority users. 
Referring to \cref{f:usecase}, 
the shared nature of CBRS band and the complexity 
of cellular network imposes new design goals:
\begin{enumerate}
    \item \textbf{Multi-band:} 
    First, since private cellular networks often comprise 
    multiple base stations (\emph{eNBs/gNBs}) operating 
    at diverse frequencies (directed by the SAS) and in close
    proximity, 
    \systemname{} requires the versatility to 
    optimize multiple frequency bands at the same time.
    
    \item \textbf{Adaptive:} Second, due to the dynamic sharing nature 
    of the CBRS band, eNB/gNB operating 
    frequencies may vary, especially when higher-tier users enter 
    the network. \systemname{} must therefore be able to adapt 
    its hardware to these changing frequencies.
    
    \item \textbf{Autonomous:}
    Finally, since \systemname{} is not a part of the existing
    cellular network infrastructure,
    it lacks direct access to both eNB/gNB 
    and the mobile user equipment (\emph{UE}). 
    Hence, its control module must function autonomously,
    without explicit 
    feedback from these 
    entities, yet remaining compliant with 
    prevailing protocols.
\end{enumerate}

Through a novel combination of hardware and software co-design
\systemname{} solves each of the foregoing design challenges.
\systemnames{} multi\hyp{}layer PCB surface design
(\S\ref{s:impl}) integrates
miniaturized, custom-designed, and tunable channel filters with
surface\hyp{}mounted amplifiers to allow the surface to 
effectively \cite{9998527}
target multiple CBRS channels simultaneously.  Integration of 
a high\hyp{}resolution LTE/5G channel monitor with a 
hardware-software real-time surface controller enables the
surface to synchronize with the Private cellular network and
adapt its hardware operation at line rate, yet remain independent
of (and thus not constrained by) the network itself.  
Thus, \systemname{} offers deployment advantages,
not requiring coordination with nor cooperation with 
both LTE/5G network equipment vendors and network operators,
realizing the first steps toward
the vision of smart surfaces for the next
generation of cellular networks \cite{9591503}.

Our experimental evaluation (\S\ref{s:eval})
measures the performance gains \systemname{}
achieves, versus a private cellular network operating
without the system.  Microbenchmarks characterize 
an average SNR gain of 8.50~dB at the physical layer 
enabled by the surface's phase shifter and filter circuitry.
Additional microbenchmarks then characterize
the ability of \systemname{} to time synchronize
to a private CBRS network operating at line rate.
End\hyp{}to\hyp{}end experiments follow, measuring
throughput for a single eNB, multiple eNBs, and the
ability of the system to adapt in real time to 
wireless channel and traffic load changes.  
Results demonstrate \systemname{} is able to achieve an 
average throughput gain of 4.36~Mbps under a single eNB, 
and 3.19~Mbps under four eNBs.

%% file: text/2-primer.tex
\section{CBRS Primer}
\label{s:primer}

\begin{figure}[th]
    \centering
    \includegraphics[width=0.99\linewidth]{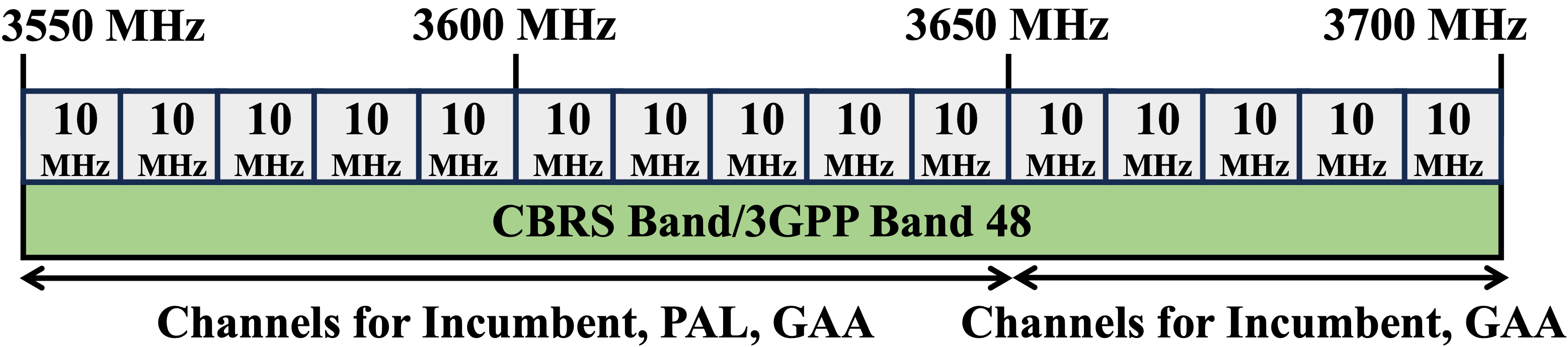}
    \caption{Frequency plan for operating in the CBRS band.}
    \label{fig:design_bandplan}
\end{figure}

The CBRS band for cellular networks is defined as Band~48 by 3GPP, 
ranging from 3550 to 3700~MHz, as shown in \cref{fig:design_bandplan}.
Although an eNB/gNB may operate anywhere within the CBRS band, it is 
typically configured with a center frequency that is a multiple of 
10~MHz, and a bandwidth of 10 or 20~MHz.

\begin{figure}[hb]
\centering
\includegraphics[width=.6\linewidth,clip]{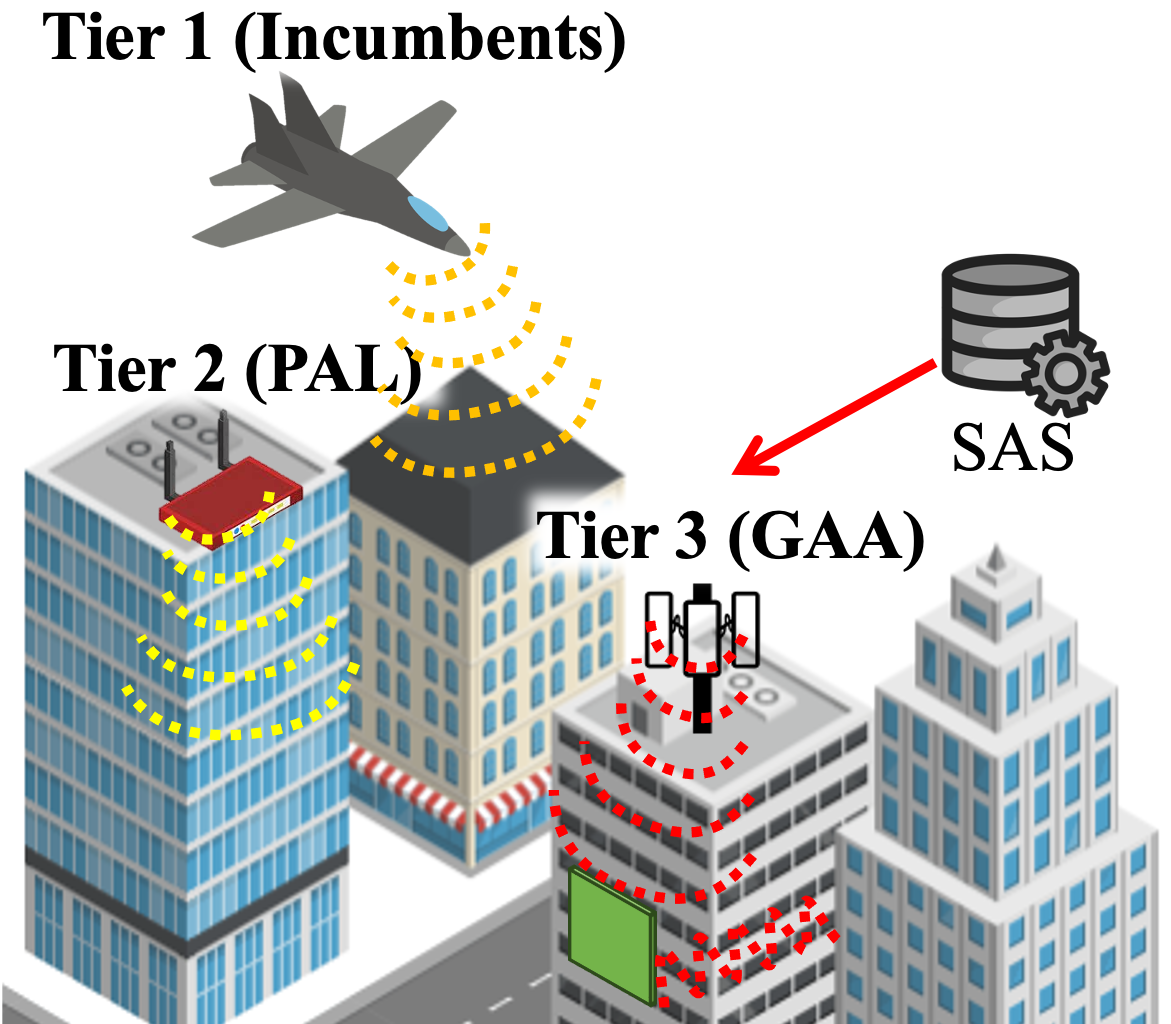}
\caption{\textbf{The CBRS band} establishes spectrum sharing rules by user
tier (incumbent, priority, and general). 
Different colors indicate different frequencies used by different user tier.}
\label{f:usecase1}
\end{figure}

The CBRS band allocates spectrum to
three user \emph{tiers} (see \cref{f:usecase1}) user system. 
The top \emph{Incumbent Access Tier}, 
prioritizes federal operations, ensuring their operations remain 
interference-free. 
In the second \emph{Priority Access License (PAL)} tier, entities 
may obtain licenses via auctions for specific geographies. 
And in the lowest, unlicensed, \emph{General Authorized Access (GAA)} tier, 
users may access the CBRS spectrum when users in higher tiers are not.
Real\hyp{}time coordination and allocation of users to these
tiers is managed by 
a \emph{Spectrum Access System (SAS)}. 
This complexity, however, 
introduces dynamism to the network, and further,
just like Wi-Fi, successful large Private LTE/5G network deployments
become the victim of their own success---the radio spectrum becomes crowded
with multiple base stations \emph{(eNBs/gNBs)} operating
concurrently, each 
operating on distinct \emph{channels} within the CBRS band,
as shown in \cref{fig:design_bandplan}.

%% file: text/3-related.tex
\section{Related Work}
\label{s:related}

A smart surface, also known as a \emph{reconfigurable
intelligent surface (RIS)}, augments---but does 
not replace---the 
wireless infrastructure of base stations and 
mobile users.  Many realizations exist, for many 
different types of wireless networks, but their 
common goal (and thus far achievement)
is to increase the performance of 
wireless links \cite{9424177, 8796365, 
visorsurf-commsmag2018, welkie2017programmable}.  
 
Earlier work in the systems and networking
community has described the design of 
smart surfaces with \emph{passive} elements
(\emph{i.e.}, elements that do not incorporate an
externally powered amplifier) to improve
the performance of Wi-Fi links: LAIA 
\cite{li2019towards}, RFocus \cite{arun2020rfocus}, and 
ScatterMIMO \cite{dunna2020scattermimo} are three
representative examples.  These systems have demonstrated
compelling performance improvements for Wi-Fi, but
do not address the added layers of complexity of
the LTE\fshyp{}5G physical, link, and medium access
control sub\hyp{}layers as summarized in \Cref{t:comparison}.

Cao \emph{et al.} design a RIS\hyp{}assisted medium access control (MAC)
protocol for \cite{cao2021reconfigurable} a Wi-Fi\hyp{}like network
using a combination of CSMA and TDMA medium access, similar to
IEEE 802.11ac Wi-Fi networks, as do Yuan 
\emph{et al.} \cite{9235486}
for multiple surfaces.  Their efforts do not delve into cellular
networks, however, nor do they implement nor experimentally evaluate
their concepts end-to-end.
\begin{table}
\footnotesize
\newcommand*\feature[1]{\ifcase#1 -\or\LEFTcircle\or\CIRCLE\fi}
\makeatletter
\newcommand*\ex[6]{#1&#2&#3&#4&#5&#6
}
\makeatother
\begin{threeparttable}
\caption{\textbf{\systemname{}:} comparison to existing surfaces.}
\label{tab:features}
\begin{tabular}{@{}lc c c c c}
\toprule
  \multirow{2}{0.7cm}{\centering Related Works}& \multirow{2}{0.7cm}{\centering Network Type} & \multirow{2}{0.5cm}{\centering Multi-band} & Layer & \multirow{2}{0.5cm}{\centering Auto-nomy} &  \multirow{2}{0.6cm}{\centering H/W Impl.}\\
\\
\midrule
\midrule
\ex{\cite{li2019towards,arun2020rfocus,dunna2020scattermimo,vmscatter-nsdi20}}{\cellcolor{red!25}Wi-Fi}{\cellcolor{red!25}Single}{\cellcolor{yellow!25}L1/L2}{\cellcolor{red!25}-}{\cellcolor{green!25}\CIRCLE}\\
\ex{\cite{chen2020pushing,10.1145/3452296.3472890}}{\cellcolor{red!25}IoT}{\cellcolor{red!25}Single}{\cellcolor{red!25}L1}{\cellcolor{red!25}-}{\cellcolor{green!25}\CIRCLE}\\
\midrule
\ex{\cite{han2021dual, rotshild2021ultra, saifullah2021dual,saeidi202122}}{\cellcolor{red!25}-}{\cellcolor{yellow!25}Dual}{\cellcolor{red!25}L1}{\cellcolor{red!25}-}{\cellcolor{yellow!25}\LEFTcircle}\\
\ex{Wall-E~\cite{cho2022towards}}{\cellcolor{red!25}Satellite}{\cellcolor{yellow!25}Dual}{\cellcolor{yellow!25}L1/L2}      {\cellcolor{red!25}-}{\cellcolor{red!25}-}\\
\ex{RF-Bouncer~\cite{li2023rf}}{\cellcolor{red!25}Wi-Fi}{\cellcolor{yellow!25}Dual}{\cellcolor{yellow!25}L1/L2}   {\cellcolor{red!25}-}{\cellcolor{green!25}\CIRCLE}\\
\ex{CrossFlit~\cite{chen2023towards}}{\cellcolor{red!25}Wi-Fi}{\cellcolor{yellow!25}Dual}{\cellcolor{yellow!25}L1/L2}        {\cellcolor{red!25}-}{\cellcolor{green!25}\CIRCLE}\\
\midrule
\ex{\textbf{WaveFlex}}{\cellcolor{green!25}LTE/5G}{\cellcolor{green!25}Quad}{\cellcolor{green!25}L1/L2/M}{\cellcolor{green!25}\CIRCLE}{\cellcolor{green!25}\CIRCLE}\\
\bottomrule
\end{tabular}
\begin{tablenotes}
\item \hfil$\feature2=\text{yes}$; $\feature1=\text{partially yes}$;
$\text{\feature0}=\text{no}$; M = \text{MAC sub-layer}
\label{t:comparison}
\end{tablenotes}    \end{threeparttable}
\end{table}

\begin{figure*}[htb]
    \centering
    \begin{minipage}[c]{1\linewidth}
        \centering
        \includegraphics[width=0.99\textwidth]{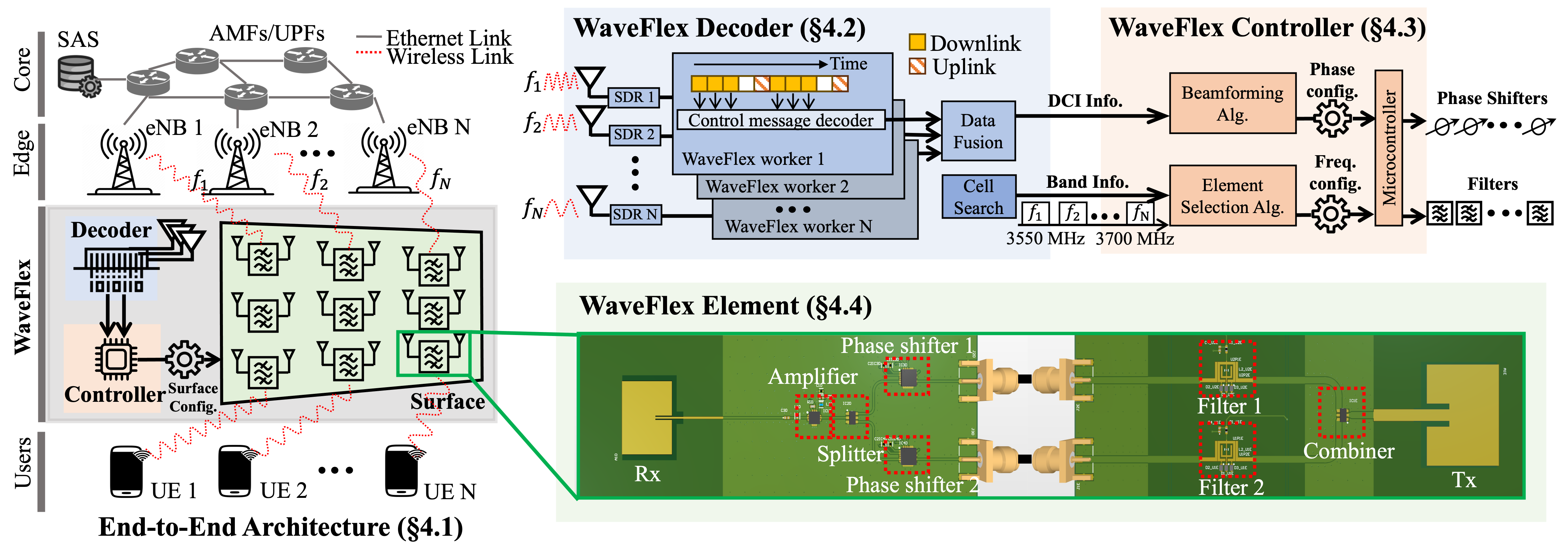}
        \caption{\textbf{\systemname{} architecture overview:} 
        The \textbf{decoder} monitors a private cellular network's 
        operation and synchronizes the system to that network,
        the \textbf{controller} optimizes the surface's configuration 
        using data from the decoder, and the \textbf{surface} 
        amplifies, filters, and modulates ambient network transmissions.}
        \label{fig:design_arch}
    \end{minipage}
\end{figure*}

Other work in both the wireless communications and computer 
systems and networking communities explores the use of 
\emph{active} elements (\emph{i.e.}, elements that do incorporate
an externally powered radio amplifier circuity). 
Long \emph{et al.} describe the architecture, compute link 
budgets, and propose novel optimization algorithms for such systems
\cite{9377648} but stop short of both 
design above the physical layer 
and a hardware realization.  LAVA
\cite{10.1145/3452296.3472890} realizes these ideas for Wi-Fi
and Zigbee networks operating in the 2.4~GHz frequency band but
stops short of integrating real-time, line-rate operation
into its implementation, and does not address Private LTE/5G
cellular networks significantly increased complexity.
Zhang \emph{et al.} use a combination of link budgets,
modeling, and experimental microbenchmarks to show that in theory, 
active surfaces can achieve multiplicative gains over passive
surfaces \cite{9998527}, yet both stop short of 
system development and integration with a full\hyp{}stack
cellular network.

Zeng \emph{et al.} \cite{10050148} investigate the use
of a smart surface
to harmonize the operation of 5G New Radio and Wi-Fi both operating
in an unlicensed (rather than shared\hyp{}spectrum) frequency band,
and propose an alternating optimization to this effect.  They
present a theoretical performance analysis, but their efforts stop 
short of both hardware and software design, and operation in shared
frequency bands.

LLAMA \cite{chen2020pushing} investigates a smart surface that
rotates signal polarization in the context of Internet of Things (IoT)
networks operating at 2.4~GHz, and backscatter reflections 
occurring from tags excited by such networks.  
VMScatter \cite{vmscatter-nsdi20} designs a MIMO 
backscatter tag and associated 
signal processing algorithms.  
While more complex than Wi-Fi \emph{per
se}, these networks are 
qualitatively simpler than the cellular networks we target here.

Dual-band smart surfaces~\cite{han2021dual, rotshild2021ultra, saifullah2021dual} have recently gained attention, but the existing designs have limitations in satisfying at least one of our criteria, namely, simultaneous multi-eNB/-user optimization, operation over more than two frequency bands, and the absence of explicit feedback from the receiver end as shown in \Cref{t:comparison}. 
Wall-E~\cite{cho2022towards} introduces a dual-band metasurface operating at two Ku bands (\textit{e.g.,} 10/15 GHz) for satellite networking, but this work stops short of hardware design and presents a theoretical performance analysis only with one user.
RF-Bouncer~\cite{li2023rf} investigates a smart surface working at two ISM bands (\textit{e.g.,} 2.4/5 GHz) to expand indoor wireless coverage. However, this work falls short of physical and link layer design and requires an explicit feedback from the receiver to control the surface.
Likewise, Chen \emph{et al.} \cite{chen2023towards} introduces a dual-band metasurface for sub-6GHz and mmWave, but it requires the signal measurement from the receiver to configure the surface. Also, its operating frequencies differ by more than $20$ GHz, which poses a less demanding design challenge than our work due to low coupling.
Saeidi \emph{et al.} \cite{saeidi202122} explores frequency-diverse leaky-wave antennas for one-shot detection of multiple wireless nodes, but this work does not integrate with the real-time system.







%% file: text/5-design.tex
\section{Design}
\label{s:design}

We begin by detailing 
\systemnames{} architecture in \cref{s:design_overview}.
\Cref{s:design_decoder} describes \systemname{} decoder, 
which enables \systemnames{} autonomy.
\Cref{s:design_control} outlines the \systemname{}
controller in the
context of a real private cellular network deployment.
\cref{s:design_hardware} describes our 
multi-channel hardware, emphasizing the tunable 
filter design and its validation 
through electromagnetic simulation. 

\subsection{System Overview}
\label{s:design_overview}

Figure~\ref{fig:design_arch} presents the architecture of \systemname{}, 
encompassing the \emph{decoder}, the \systemname{} 
\emph{controller}, and our frequency\hyp{}tunable 
multi\hyp{}band \emph{surface} itself.

\parahead{\systemname{} decoder}
A distinctive feature of \systemname{} is its capability to operate 
independently without needing explicit information from the SAS, 
\emph{User Plane Function (UPF)}, or Access and Mobility Management 
Function (AMF) in the core.  
Instead, \systemname{} monitors these components of the
cellular network by wirelessly monitoring 
the control plane traffic between the eNB and UE.
Keeping in step with the time schedule of the cellular network,
the \systemname{} decoder synchronizes with the eNB, and decodes granular, 
millisecond\hyp{}level \emph{Downlink Control Information (DCI)}.
The decoded DCI is sent to the \systemname{} controller for subsequent 
processing.

\parahead{\systemname{} controller}
Based on the received DCI, the \systemname{} controller 
runs a beamforming algorithm and an element selection algorithm to calculate
appropriate phase and filter configurations, respectively, which are subsequently 
sent to the microcontrollers for real-time adjustments.

\parahead{\systemname{} surface}
The \systemname{} surface (\Cref{fig:design_arch})
consists of a series of amplified unit elements, 
the circuitry of which is illustrated in 
the lower right of the figure.  Each unit
element operates indpendently on each of
two CBRS channels (\emph{cf.}~\S\ref{s:primer}), as
described in \cref{s:design_hardware}.
An integrated microcontroller applies phase and 
filter configurations to phase shifters and 
custom tunable filters, 
as instructed by the \systemname{} Controller. 

\input{text/5-design_decoder}

\input{text/5-design_controller}
\input{text/5-design_hardware}

\input{text/link_budget}

%% file: text/5-design_decoder.tex
\subsection{Decoder Design}
\label{s:design_decoder}

The \systemname{} decoder performs two functions. It first searches for 
all nearby private LTE eNBs within CBRS band. Subsequently, the decoder 
synchronizes with the detected eNBs, decode their control channel to 
extract DCI. The list of eNBs and extracted DCI is then relayed to
the \systemname{} controller for filter control and beamforming phase 
control, respectively.

\parahead{Cell search}
The \systemname{} decoder executes cell search to identify eNBs within 
CBRS band, which involves traversing the CBRS band and attempting to 
detect LTE synchronization signals and decode broadcast channels. 
Given this, the \systemname{} decoder employs a step size of 10~MHz for 
cell search.
If the decoder loses an eNB's signal due to eNB operating frequency changes, 
it promptly reinitiates the cell search to reacquire the eNB's frequency.

\parahead{LTE TDD control channel decoder}
Prior works \cite{xie2022ng, kumar2014lte} have demonstrated the 
feasibility of an external entity decoding the control channel to 
extract DCI. 
We design a decoder tailored for Time Division Duplex (TDD), 
which is the duplexing mode of CBRS band private LTE network.
TDD operates by segmenting time into periods, where some periods 
dedicated for downlink data transmissions and others for uplink.

\begin{figure}[th]
    \centering
    \includegraphics[width=0.99\linewidth]{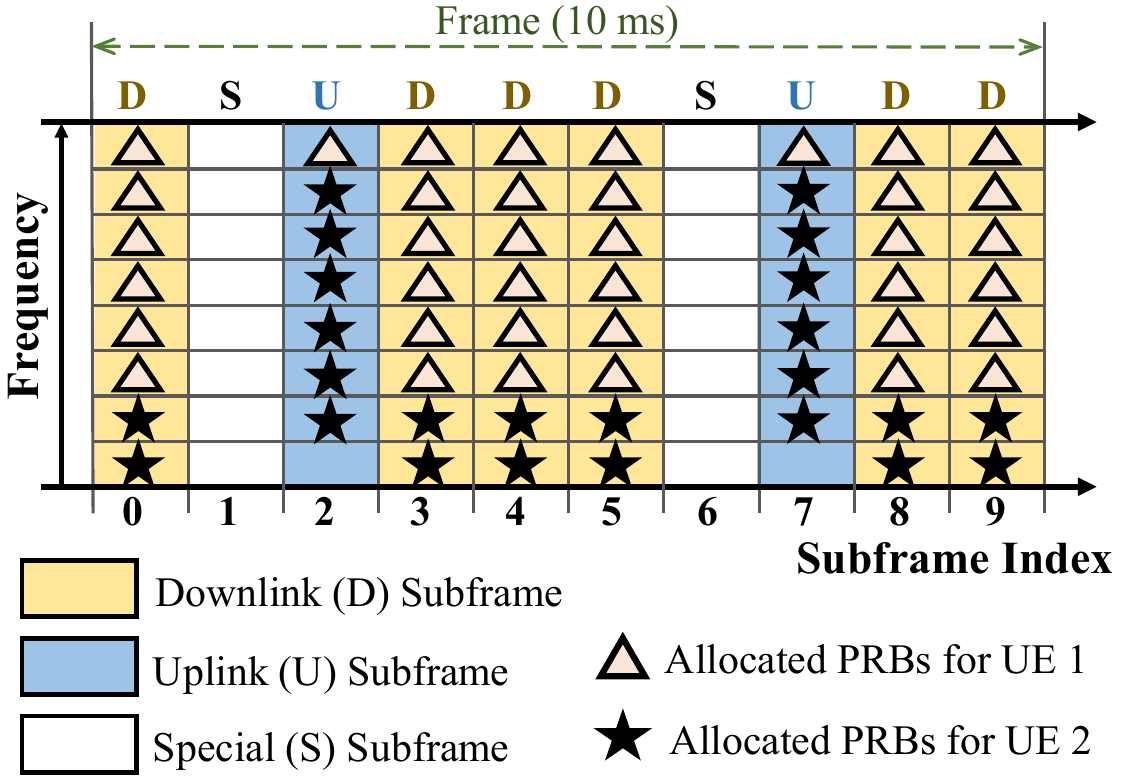}
    \caption{\textbf{CBRS frame structure configuration},
    which reflects time-division duplex operation, designating
    certain \emph{sub\hyp{}frames} within a 10\hyp{}millisecond frame to carry
    downlink (D) or uplink (U) data traffic to 
    one or more UEs.  Special (S) subframes contain reference information
    for PHY operation.}
    \label{fig:design_TDDFrame}
\end{figure}

To decode the TDD control channel, we first determine 
the arrangement of uplink (U), downlink (D), and 
special (S)\footnote{S subframes
contain downlink and uplink pilot time slots for reference signals, 
guard periods for tolerance to synchronization uncertainty.}
subframes within a frame.
We show an example \emph{frame structure configuration} of LTE TDD in 
Figure~\ref{fig:design_TDDFrame}.\footnote{We show one of seven possible 
frame structure configurations (the configuration of our CBRS 
eNB)---our system generalizes to others.} 
Within each frame, the network shifts from downlink to uplink in subframes 
two and seven. 

The frame structure configuration information is carried in 
\emph{System Information Block 1 (SIB1)}, which occurs every 20~ms and 
is broadcasted in plain text.
The \systemname{} decoder first decodes SIB1 to acquire the frame structure 
configuration, and based on this information, decodes the control 
channel for downlink subframes and special subframes, skipping uplink subframes, 
since DCI is located only in downlink and special subframes.
In addition to message decoding, the \systemname{} decoder also computes 
per-subframe \emph{Reference Signal Received Power (RSRP)}
by utilizing the reference signal transmitted within each subframe, 
regardless of whether there is data transmission within that subframe or not.

%% file: text/5-design_controller.tex
\subsection{Controller Design}
\label{s:design_control}

Using the \systemname{} decoder, the \systemname{} 
controller calculates optimal phase 
shifter and filter configurations of each element 
to maximize network throughput.




\subsubsection{Traffic Scheduling in TDD CBRS}
\label{s:design_TDD}

The TDD nature of CBRS networks imposes challenges
on the design of the \systemname{} controller.
First, while downlink traffic demand
typically exceeds uplink demand by a ratio of 2--3$\times$ 
\cite{yang-iet16}, there is variation between UEs and in time,
and so uplink (as well as downlink) traffic must 
be made reliable in order to facilitate the flow 
of downlink traffic.
Second, when multiple UEs are active,
the RAN will schedule different UEs in different
subframes, and so
\systemname{} must apply different phase shifts
to optimize different subframes, at millisecond
time granularities. 
Referring again to the CBRS frame structure
(\cref{fig:design_TDDFrame})
the eNB switches from downlink to uplink in subframes two and seven.
The controller needs to synchronize with the eNB
and modulate the 
phase shifts that it applies to the surface in order 
to keep pace with the CBRS frame structure
link direction.


\begin{figure}[htb]
    \centering
    \includegraphics[width=0.99\linewidth]{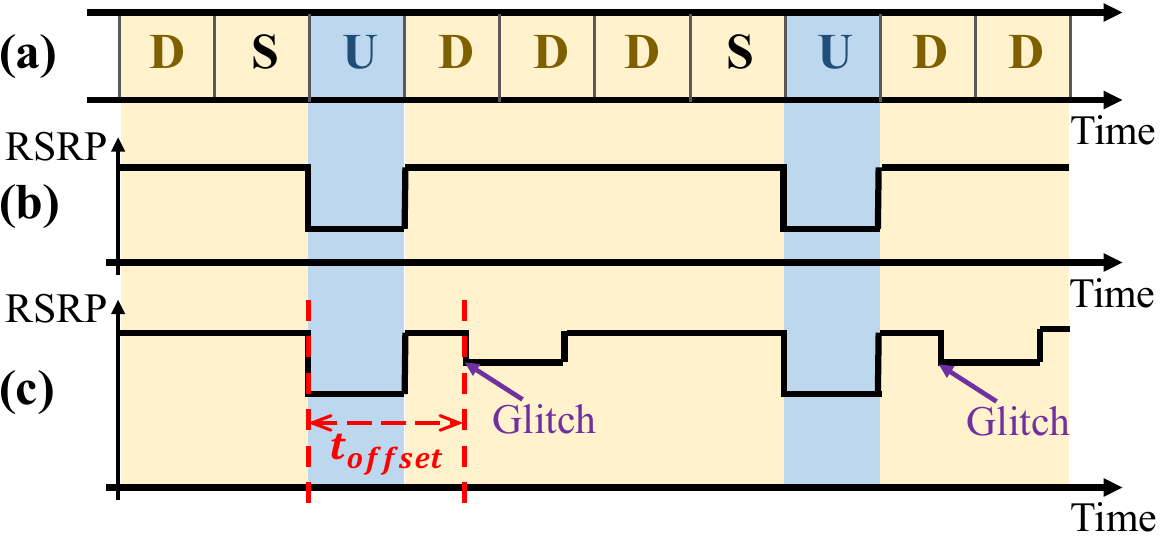}
    \caption{\textbf{Operation of \systemnames{} time synchronization} algorithm:
    \textbf{(a)} frame structure; \textbf{(b)} synchronized state;
    \textbf{(c)} unsynchronized state.}
    \label{fig:design_TDDToggle}
\end{figure} 

The time-switched nature of TDD places a precise 
synchronization requirement between the \systemname{} controller and the eNB. 
The primary objective is to align the switching of phase 
configurations in each subframe with 
the respective downlink and uplink traffic schedules.
To address the synchronization problem, our strategy centers on 
monitoring the periodic pattern of
RSRP changes, as shown in \cref{fig:design_TDDToggle}. 
When the \systemname{} controller's element switching pattern
is synchronized with the LTE TDD schedule 
(\cref{fig:design_TDDToggle}(a)), 
we observe a periodicity matched to the TDD schedule
as the active transmitter (eNB or any UE) 
in the schedule changes, as shown in 
\cref{fig:design_TDDToggle}(b).

When the \systemname{} controller's element switching pattern
is instead offset by an amount $t_{\mathrm{offset}}$ from the
LTE TDD schedule, we instead observe brief changes in 
the RSRP curve, as shown in Figure~\ref{fig:design_TDDToggle}(c), 
where the labeled \emph{glitches} are caused by the mismatch between 
the downlink\fshyp{}uplink phase configuration 
transitions of the surface and the (correct) TDD schedule of the eNB.
By analyzing the positions of the two periodic RSSI changes, the 
\systemname{} controller estimates $t_{\mathrm{offset}}$, 
and adjusts its schedule by $-t_{\mathrm{offset}}$ to converge 
towards alignment with the network's TDD schedule.

\subsubsection{Blind Beamforming Algorithm}
\label{s:design_BF_straw}

To introduce the \systemname{} beamforming
algorithm, we start with the description of 
the algorithm under ideal conditions with only a single UE and a single eNB.
We then extend our algorithm to cover multiple UEs and multiple eNBs. 

Our algorithm uses \emph{blind-beamforming},
to run without \emph{Channel State Information (CSI)},
\cite{tseng2011bio, tseng2014robust, jayaprakasam2017distributed}.
Suppose we have $K$ elements in the smart surface, and the 
phases of the $K$ phase shifters are 
\begin{equation}
    \theta = [\theta_1, \theta_2, \cdots \theta_K]^T.
\end{equation}
The blind beamforming algorithm searches for the 
optimal $\theta^*$ that maximizes the 
the channel condition:
\begin{equation}
    \theta^* = \argmax_\theta M(\theta)
    \label{eqn:max_m}
\end{equation}
where $M$ (defined below) characterizes the quality of the channel 
between the transceivers.
Blind beamforming solves the problem via an iterative search:
it applies a random perturbation $\delta[n]$ at the $n^{\mathrm th}$ 
iteration:
\begin{equation}
\theta[n] = \theta[n-1] + \delta[n], 
\end{equation}
where $\theta[n]$ are the phase settings in the $n^{\mathrm th}$ iteration,
repeating the above process until convergence.

\parahead{Characterizing cellular channel quality}
\systemnames{} controller works independently without direct feedback from eNB or UE.
Consequently, it cannot directly query the UE for standard channel quality 
metrics such as the reference received signal power (RSRP), 
signal to noise ratio (SNR), or CSI.
However, we observe that the eNBs implement a bitrate adaption algorithm that 
adjusts the modulation and coding rate
to cope with the quality of the channel between the UE and the base stations,
making this rate an excellent proxy for assessing cellular channel conditions.
More importantly, the eNB broadcasts the rate index 
and it is thus decodable by the \systemname{} decoder.
The \systemname{} decoder serves as a sniffer to decode the bit rate
the eNB broadcasts,
and calculates the physical data rate $R_w$, 
an estimate of the greatest number of bits that can be transmitted over 
one physical resource block
without causing errors. 

Given the inherent variability of the estimated data rate, 
we adopt an average rate 
$\overline{R_w}$ spanning a preset number of subframes to mitigate the 
influence of bit rate index fluctuations:
\begin{equation}
\overline{R_w} = \frac{\sum_{i=1}^{N_{sf}}R_w^i}{N_{sf}},
\label{eqn:BF_avgRate}
\end{equation}
where $R_w^i$ denotes the rate of the $i^{th}$ subframe, and $N_{sf}$ 
represents the number of subframes used for the averaging process.
We use this averaged rate $\overline{R_w}$ as the optimization target 
for a single UE scenario:
\begin{equation}
    M_{single}(\theta) = \overline{R_w}(\theta).
    \label{eqn:m_single}
\end{equation}
Combing \cref{eqn:max_m,eqn:m_single}, 
we search for the optimal $\theta$ that maximizes 
channel quality $M_{single}(\theta)$.

\parahead{Extension to multiple UEs}
The blind-beamforming algorithm simultaneously optimizes 
the channel conditions for multiple UEs.
To ensure fairness among UEs, we introduce a channel condition metric 
tailored for the multi-UE context, 
which is expressed as:
\begin{equation}
M_{multi} = \frac{\sum_{j=1}^{N_{UE}}\overline{R_w(j)}\times N_{PRB}(j)}{\sum_{j=1}^{N_{UE}} N_{PRB}(j)},
\label{eqn:multi_m}
\end{equation}
where $\overline{R_w(j)}$ represents the averaged rate of the $j^{th}$ UE, 
and $N_{PRB}(j)$ corresponds to the summed number of PRBs 
allocated to the $j^{th}$ UE in $N_{sf}$ consecutive subframes.


\parahead{Mobility}
As illustrated in Equation~\ref{eqn:max_m}, the iterative search process 
maintains a historical record of the highest channel quality value, 
denoted as $M_{MAX}$. This value serves as a benchmark for iteratively 
searching phase configurations to optimize the channel condition. 
However, in a dynamic environment, $M_{MAX}$ expires when wireless channel changes, 
necessitating a mechanism to detect and react to such channel changes.

We use consecutive negative feedbacks as a sign of possible channel change.
Once the number of consecutive negative feedbacks achieves a pre-defined 
threshold, $N_{nf}$, \systemname{} reconfigures phase settings that corresponds
to $M_{MAX}$, and then measure the channel quality again, which obtains 
$M_{new}$.
If $M_{new}$ is smaller than $M_{MAX}$, \systemname{} identifies 
that a channel change event happens, and then resets its beamforming parameters, 
including $M_{MAX}$, and $\theta[n]$. 
Our experimental results in \cref{s:eval_mobile} demonstrate that this mechanism
is able to cope with changes in the wireless channel, even in mobile cases.

\subsubsection{Multiple-eNB element selection algorithm}

The two-way element design naturally fits in with the two-eNB scenario.
For scenarios involving more eNBs, we use a greedy search based 
element selection algorithm to assign elements to the eNBs. 
Initially, to ensure fairness among eNBs, we evenly and randomly assign the 
$K$ elements to the $N_{eNB}$ eNBs and start running the beamforming algorithm.
After the beamforming converges, we record the channel quality $M$.
We then perturb the element selection by choosing a handful of elements to 
alternate eNBs while keeping the number of elements assigned to each eNB 
unchanged.  Following another beamforming convergence, the updated channel 
metric $M$ is compared against its predecessor. 
Favorable allocations are retained, while detrimental ones are discarded.

This iterative procedure is maintained until a stable solution is achieved or 
until there's a shift in the eNB operating frequencies. To accommodate the 
dynamism in eNB frequencies, the greedy search is reinitialized based on the 
updated eNB count $N_{eNB}$.

%% file: text/5-design_hardware.tex
\subsection{Hardware Design}
\label{s:design_hardware}
\label{s:design_filter}

\begin{figure}[htb]
        \centering
        \includegraphics[width=0.85\linewidth]{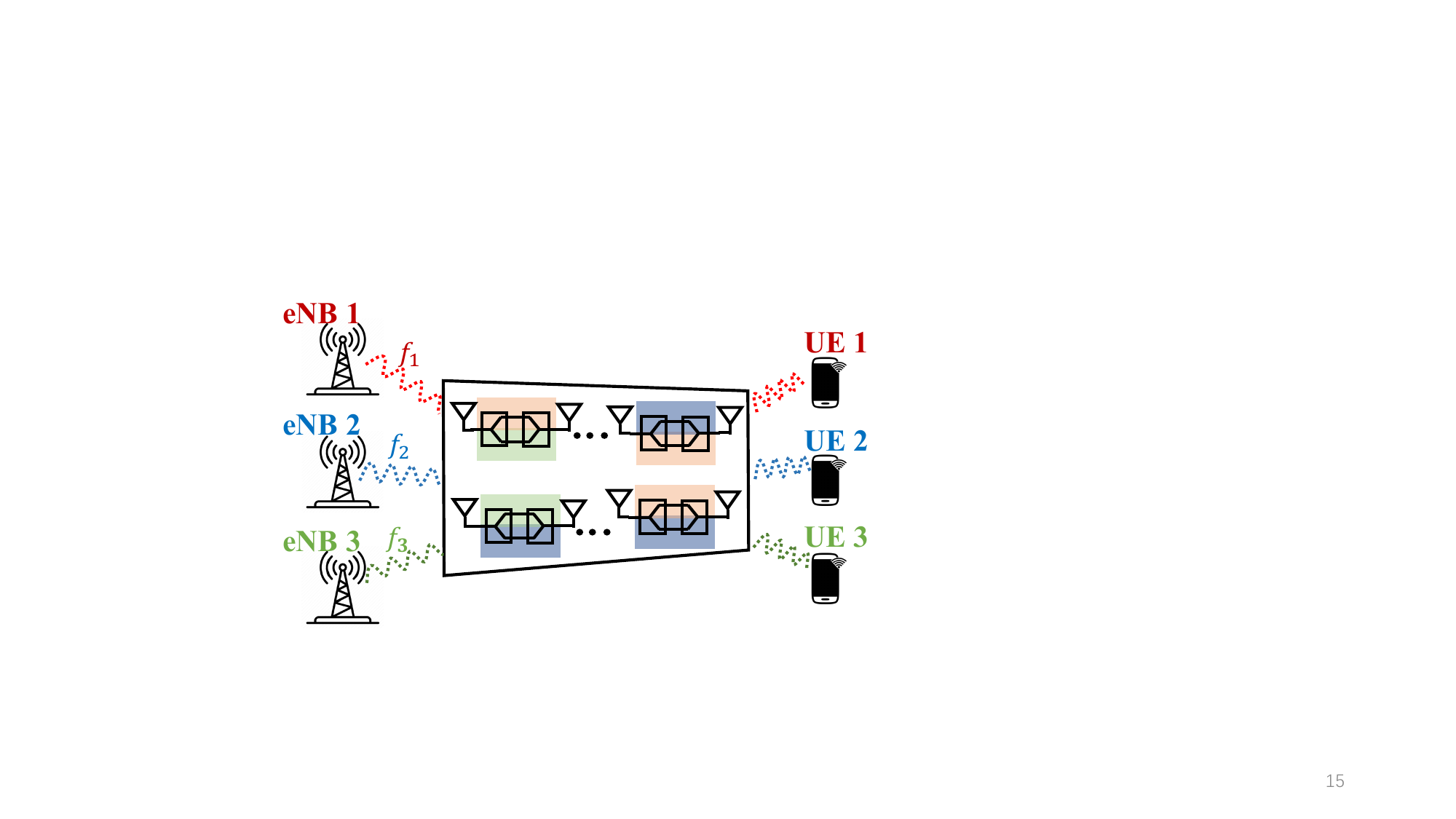}
        \caption{\textbf{Multi-channel operation goal:} different elements affect different CBRS channels independently.}
        \label{fig:design_multichannel}
\end{figure}

CBRS networks' design shapes the 
goals of \systemname{}'s hardware design, which we
illustrate in \cref{fig:design_multichannel}:

\begin{enumerate}
    \item \textbf{Frequency tunability:}
    Private CBRS networks generally involve 
    multiple eNBs working on different \emph{channels}, \ie 
    over non-overlapping frequency ranges.
    To cope with the dynamics in those operating channels, 
    \systemnames{} hardware must be flexible to adjust its operating frequency.
    
    \item \textbf{Concurrent cross-channel isolation:}
    Further, in a multi\hyp{}eNB network with concurrent, unsynchronized
    TDD operation at each eNB, \systemnames{} hardware must 
    help each channel, while not hindering others.
\end{enumerate}

\begin{figure}
    \centering
    \includegraphics[width=.95\linewidth]{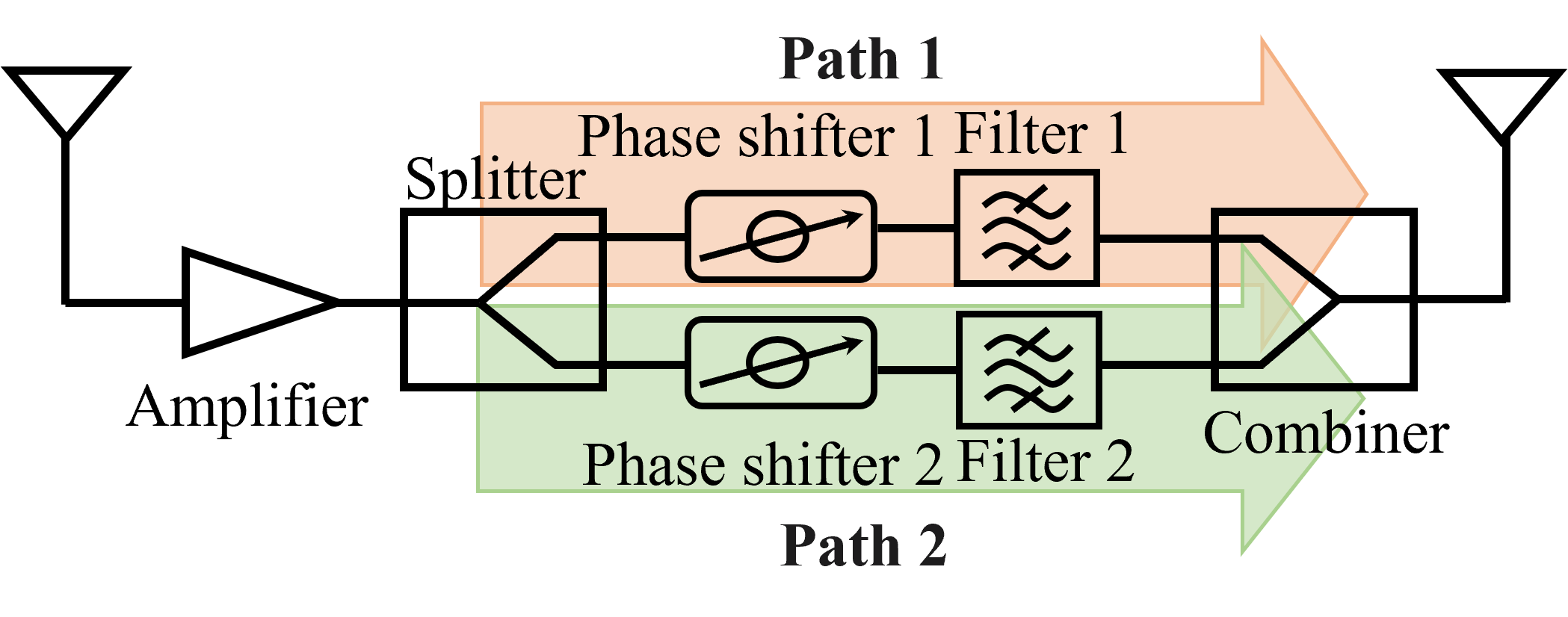}
    \caption{\textbf{Circuit schematic} of the \systemname{}
    unit element.}
    \label{fig:element_schematic}
\end{figure}

The \systemname{} \emph{unit element} 
consists of two \emph{patch antennas} that receive and
transmit signal respectively, one \emph{phase shifter} to control 
and adjust the transmitted signal phase,
an \emph{amplifier}
to boost the output signal level,
and one \emph{pass-band filter} to achieve our goals of
tunability and isolation,
as illustrated in \cref{fig:element_schematic}. 

To facilitate multi-channel operation, we integrate 
two single-channel elements 
into a unified multi-channel element, as illustrated
in the figure.
A two-way power splitter-combiner 
divides the received signal and 
directs each copy into a respective phase shifter and filter.
The signals produced by the two filters are 
subsequently recombined using another splitter-combiner.
Such a configuration enables each single-band unit element to independently 
modulate the signal phase and tune 
the two respective operating frequencies of each band-pass filter.
Our steep roll-off microstrip filter 
minimizes cross-talk between the two signal paths 
residing on the same board. 

\subsubsection{\systemnames{} filter design}
Our filter design goals are as follows:
\textbf{(1)}~We require a \emph{bandpass} filter that, for each 
of the two signal paths traversing the unit element,
allows the frequency range of a single CBRS channel
that the controller designates
(the \emph{pass-band}) to pass
through that path with minimal loss, while blocking to 
a large degree frequencies 
even minimally outside of that
same range.  This allows the path to help the channel of interest
without negatively impacting other, possibly adjacent CBRS channels
(see \S\ref{s:primer} on p.~\pageref{s:primer}).
\textbf{(2)}~Our filter must permit agile adjustment of
the pass-band 
to accommodate the dynamic nature of the CBRS frame
structure (see \cref{fig:design_TDDFrame} on 
p.~\pageref{fig:design_TDDFrame}). 
\textbf{(3)}~We require a filter 
that is cost-effective so that the \systemname{} surface
can scale up to large unit element counts,
for a realistic deployment.

\begin{figure*}
    \centering
    \begin{subfigure}{0.20\linewidth}
        \centering
        \includegraphics[width=0.8\linewidth]{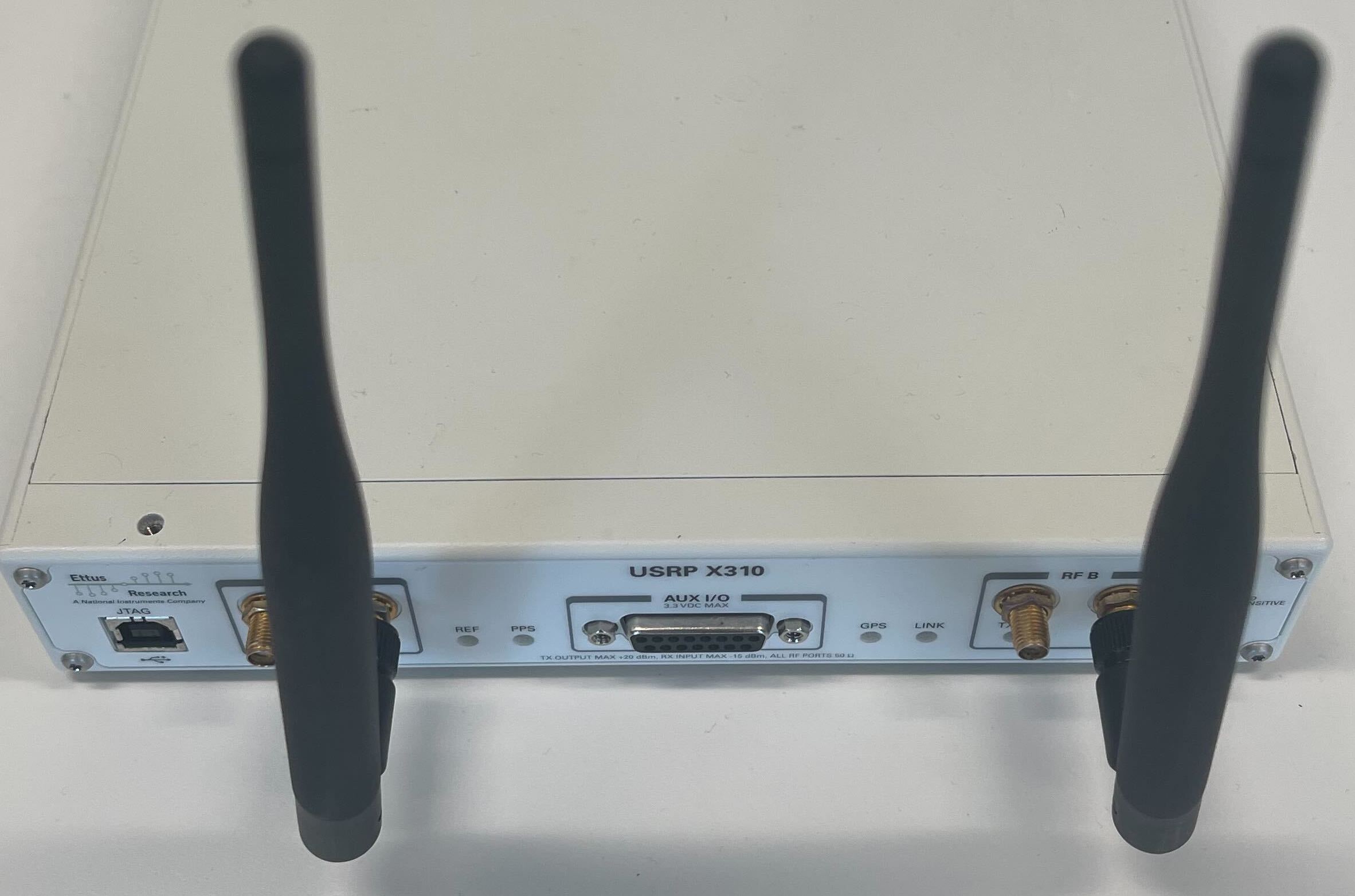}\\
        \includegraphics[width=0.8\linewidth]{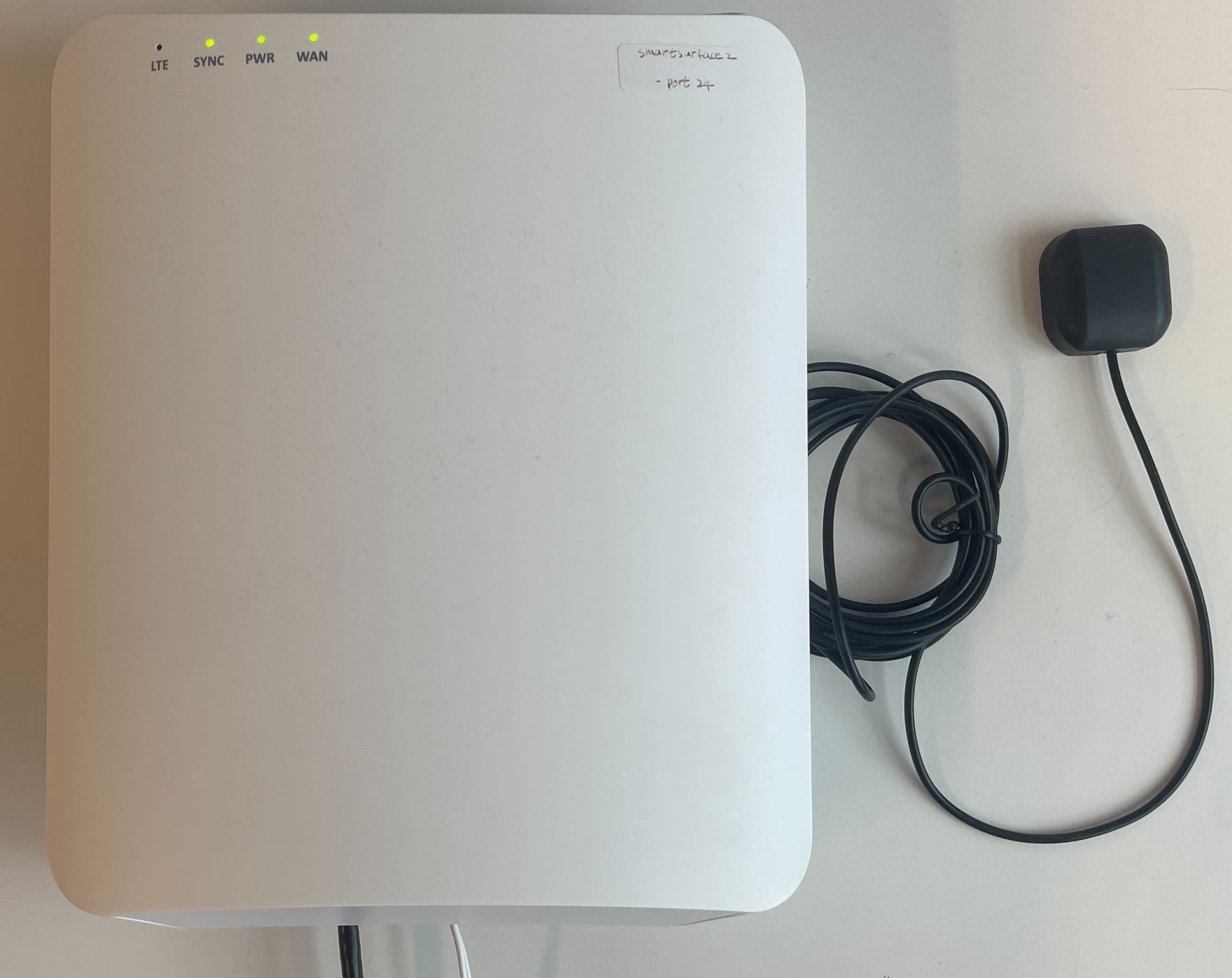}
        \caption{USRP X310 \emph{(upper)}; Sercomm cell \emph{(lower)}.}
        \label{fig:hw_impl:usrp_cell}
    \end{subfigure}
    \hfill
    \begin{subfigure}{0.20\linewidth}
        \includegraphics[width=\linewidth]{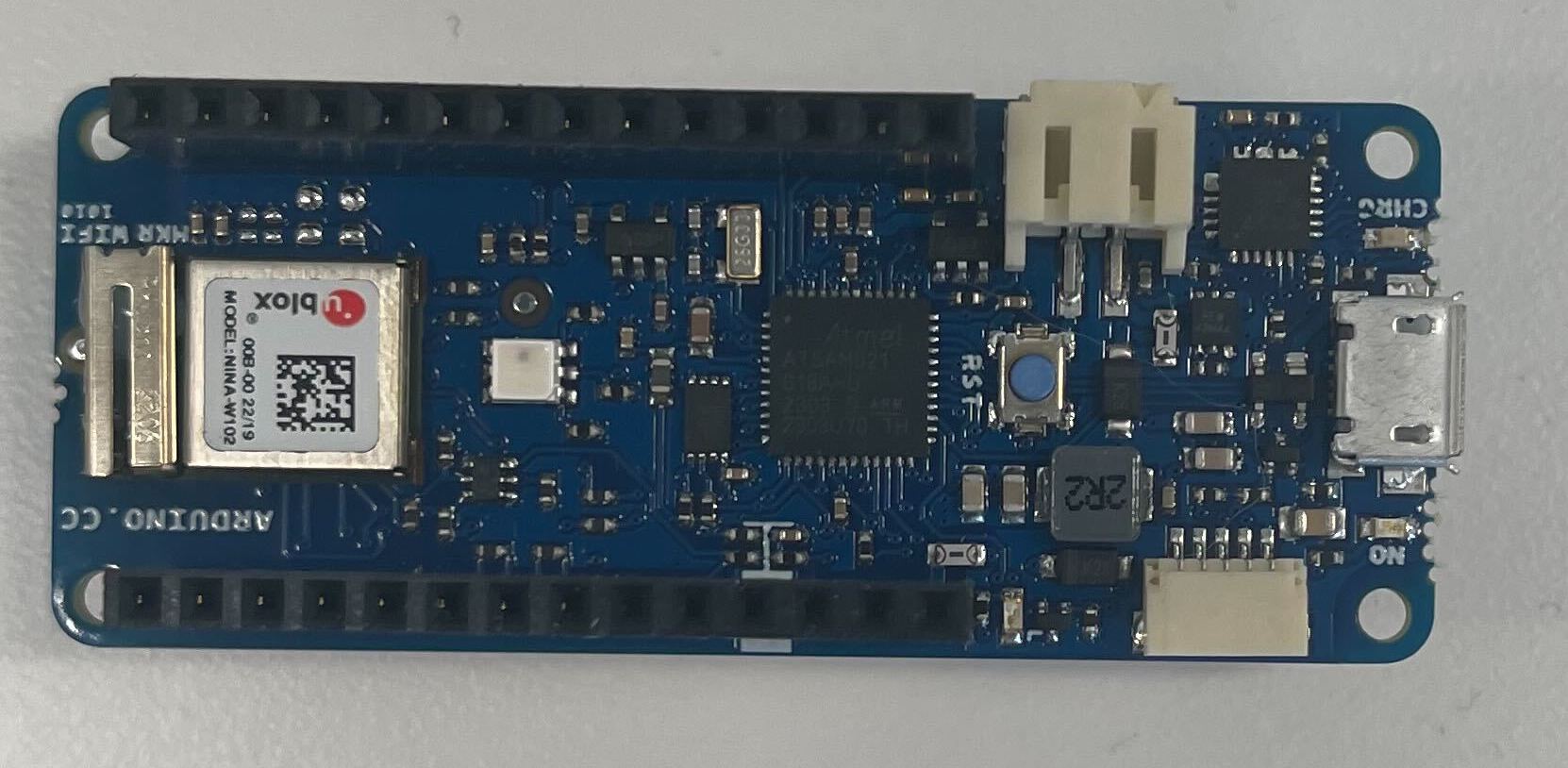}
        \includegraphics[width=\linewidth]{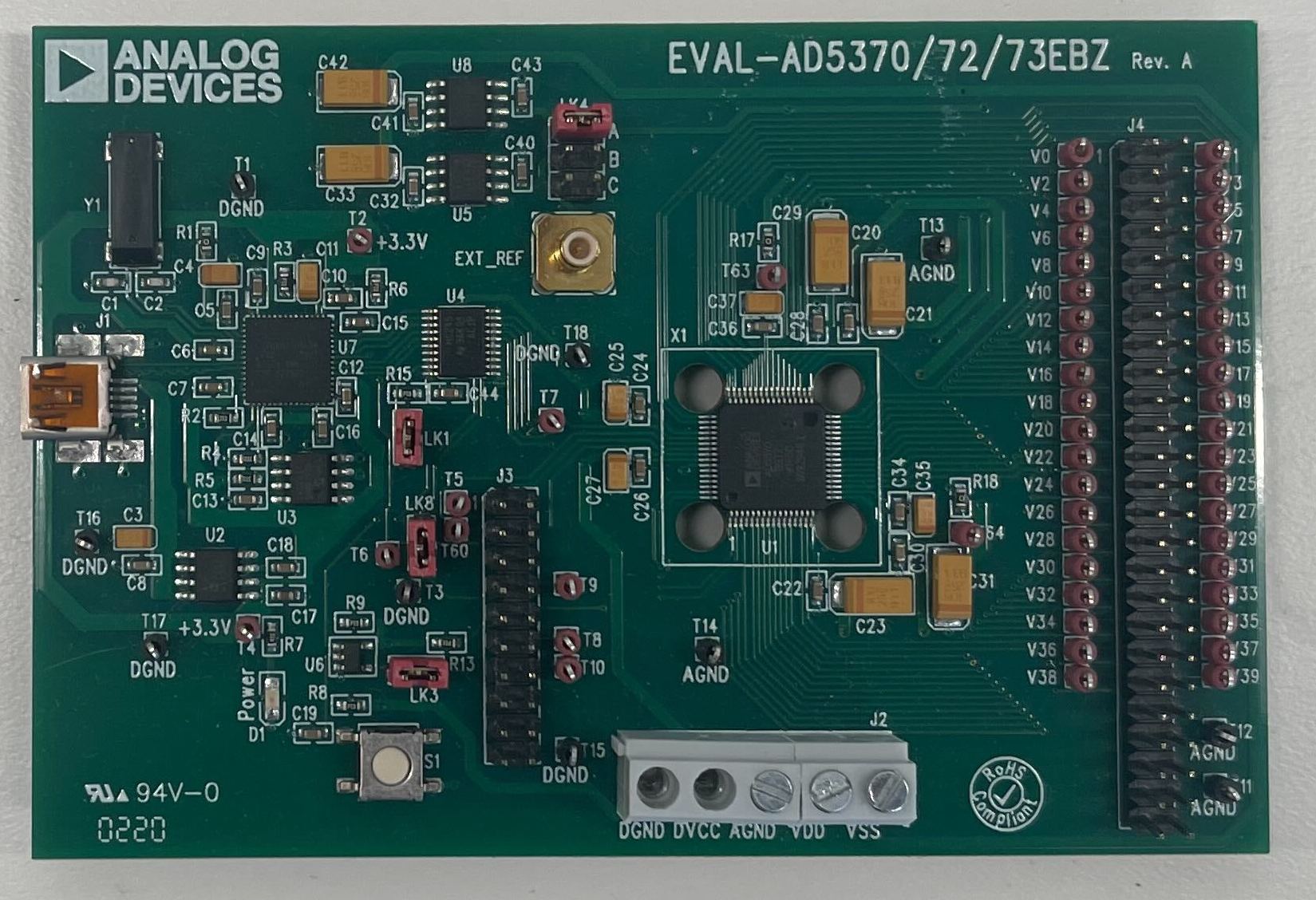}
        \caption{Arduino controller \emph{(upper)}; DAC \emph{(lower)}.}
        \label{fig:hw_impl:arduino_dac}
    \end{subfigure}
    \hfill
    \begin{subfigure}{0.2\linewidth}
        \includegraphics[width=\linewidth]{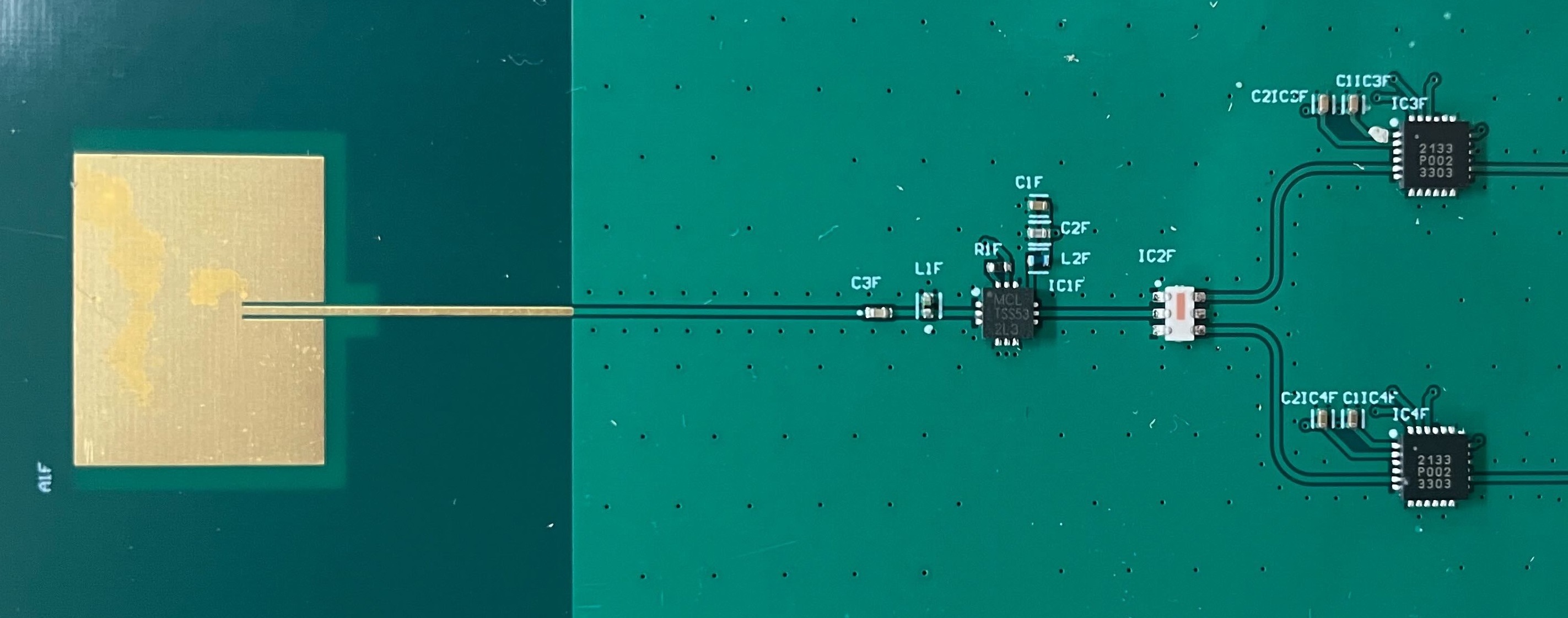}\\\vspace*{2ex}
        \includegraphics[width=\linewidth]{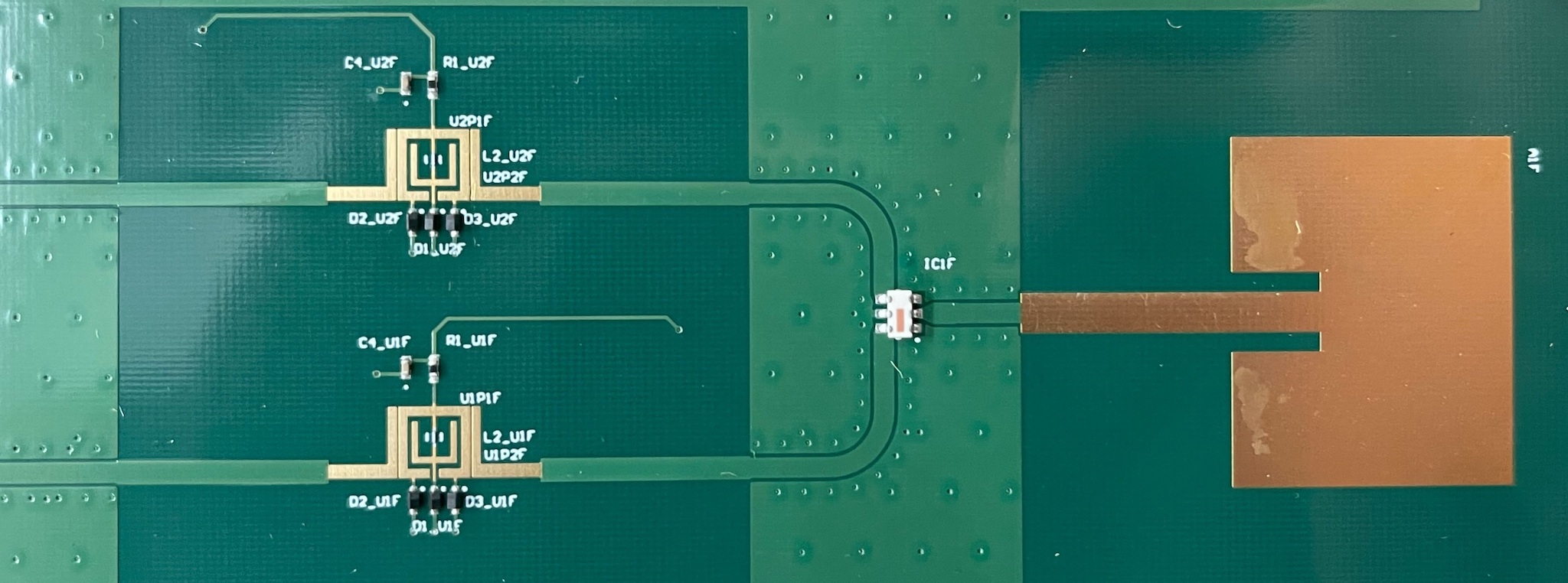}
        \caption{Receive-half \emph{(upper)}; transmit-half \emph{(lower)} custom PCB unit elements.}
        \label{fig:hw_impl:custom_pcb}
    \end{subfigure}
    \hfill
    \begin{subfigure}{0.3\linewidth}
        \includegraphics[width=\linewidth,trim={0 1cm 0 3cm},clip]{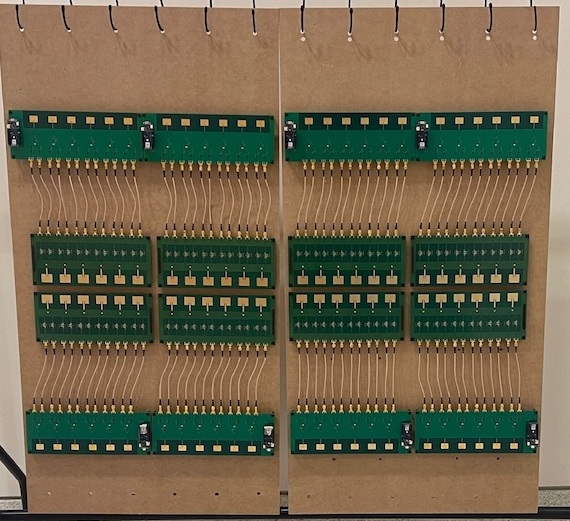}
        \caption{The integrated prototype 48-element, one sq. meter
        \systemname{} surface deployed on a temporary wall.}
        \label{fig:hw_impl:surface}
    \end{subfigure}
    \caption{\textbf{Hardware implementation:} \systemname{} combines commodity
    Private LTE and electronics hardware with custom multi-layer PCBs design into
    an integrated system.}
    \label{fig:hw_impl_container}
\end{figure*}

We use dual-mode microstrip filter \cite{hong2004microstrip, hong2007dual} 
to design our band-pass filter, which offers advantages such as space 
conservation, and cost-effectiveness.
To meet our requirements of arbitrarily tuning our operating frequencies, 
we augment the dual-mode microstrip filter with 
varactors~\cite{hong2004microstrip, tang2008compact}.
The varactor is a voltage-dependent capacitor, by applying voltage on it, 
we are able to change its capacitance, which affects the filter's 
operating frequency.\footnote{This relationship stems from how electrical 
signals interact with capacitance: capacitance determines how quickly the 
filter reacts to incoming signals, and this reaction speed correlates 
directly with the filter's resonant frequency.} 
We detail the design of the tunable filter and the derivation of 
its dimension parameters in \cref{appen:microfilter} and \cref{appen:filter}.

%% file: text/link_budget.tex
\begin{figure}
    \centering
    \includegraphics[width=.65\linewidth]{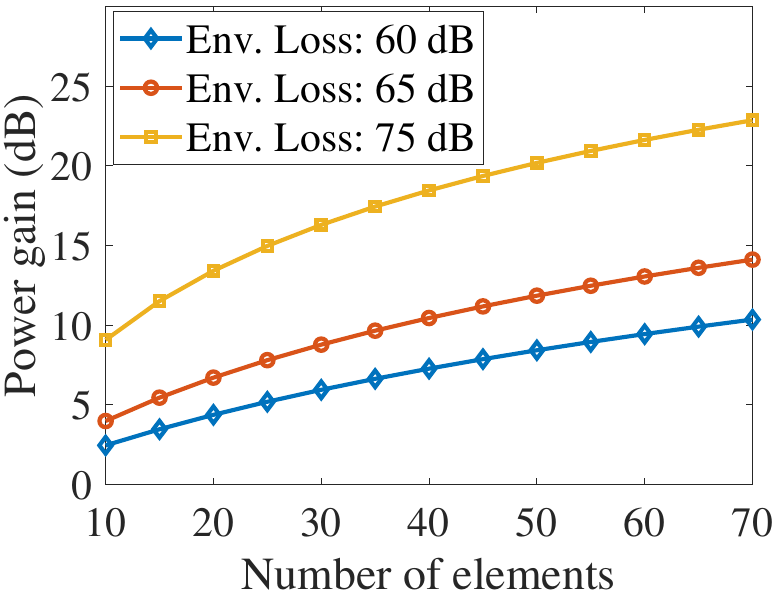}
    \caption{\textbf{The received signal power gain in theory:} the curve is obtained by adjusting number of elements and environment signal loss.}
    \label{fig:link_budget}
\end{figure}

\subsubsection{Link Budget}
\label{s:link_budget}

To understand the fundamentals of \systemnames{} hardware effectiveness and
place our later empirical results into context,
we derive a link budget of a CBRS network using \systemname{}.
When a signal transmitted by the base station travels distance $d_b$ to reach the \systemname{} element,
the signal loss it experiences can be calculated according to the free space path loss:
\begin{equation}
   L_{be} = \left( \frac{4\pi d_b}{\lambda} \right)^2
\end{equation}
where the $\lambda$ is the wavelength of the signal.
The signal then be received by the receiving antenna,
going through all the physical components on the element,
and then transmitted again by the transmitting antenna 
of the element.
We characterize the signal gain $G_{ele}$ after going through one element as:
\begin{align}
     G_{ele} = \ &G_{rx} + G_{tx} + G_{amp} - L_{split} - L_{comb} \\ \nonumber
               &- L_{phase} - L_{filter} - L_{line} - 3
\end{align}
where $G_{rx} = 2.46$ dB, $G_{tx} = 2.46$ dB, and $G_{amp}=16.65$ dB are the gain of the transmitting antenna, receiving antenna
and the power amplifier.
The $L_{split}=-0.64$ dB, $L_{comb} = -0.64$ dB, $L_{phase} = -2.5$ dB, $L_{filter} = -5$ dB, and $L_{line} = -1$ dB represent the 
loss introduced by the signal splitter, signal combiner, phase shifter, the filter and the transmission line
of the element. The three dB loss represents the signal loss introduced by the splitting operation.
Therefore, the signal emitted by the element undergoes an additional signal loss before reaching the UE. 
This signal loss is determined by the distance between the UE and the element $d_u$ and can be calculated as follows:
\begin{equation}
   L_{eu} = \left( \frac{4\pi d_u}{\lambda} \right)^2
\end{equation}
Therefore, the total loss of a signal that travels through the \systemname{} element before reaching the UE is:
\begin{equation}
    L_{ele} = G_{ele} - L_{be}  - L{eu}.
    \label{eqn:ele_lost}
\end{equation}
Assuming the transmission power of the base station is $P_b$, the power loss of the environment path is $L_{env}$, 
and there are $K$ \systemname{} elements in the smart surfaces,
after perfectly adding the signal of the environment path and the signals from all $K$ elements,
the gain $G_s$ of such a smart surface is given as:
\begin{equation}
    G_s = \textit{db} \left( \frac{  N \cdot \textit{db}^{-1}(P_b - L_{ele}))}{  \textit{db}^{-1}(P_b - L_{env})} \right)
\end{equation}
where the operation $\textit{db}(\cdot)$ represent transforming the amplitude into power in dB
and the operation $\textit{db}^{-1}(\cdot)$ represents the reverse operation of $\textit{db}(\cdot)$ .
We have depicted the achieved gain as a function of the number of WaveFlex elements in \cref{fig:link_budget}. 
It is evident from the plot that the signal gain exhibits a linear increase with the number of elements increases. 
Moreover, when the environmental path experiences greater signal loss, 
WaveFlex demonstrates its ability to substantially amplify the signal strength, resulting in even higher gains.

%% file: text/6-impl.tex
\section{Implementation}
\label{s:impl}

\parahead{Surface}
Our \systemname{} surface integrates 
components on custom multi-layer PCBs, as shown in 
Figure~\ref{fig:hw_impl_container}. 
The surface has 48 elements in total, 
with a surface area of $119 \times 76$~cm$^2$.
We use SCN-2-35+ splitter-combiner \cite{splitter} 
and TSS-53LNB3+ low noise bypass amplifier \cite{amplifier} from 
Mini-Circuits in our unit elements (\cref{fig:hw_impl:custom_pcb}).
We employ Macom MAPS-010144 four-bit phase shifters \cite{phaseShifter}
and an Arduino MKR Wi-Fi 1010 \cite{arduino} to control the phase of 
\systemname{} elements at a granularity 
of $\pi/8$ (\cref{fig:hw_impl:arduino_dac}).
Our tunable filter is depicted in \Cref{fig:hw_impl:custom_pcb} \emph{(lower)}. 
For tunable filter control, we use 40-channel AD5370 DACs from Analog 
Devices\cite{dac} to apply variable bias voltages ranging from 2 to 6~V 
to Macom MAVR-011005-12790T varactors \cite{varactor}.

\parahead{Decoder}  The \systemname{} decoder is modified
from NG-Scope \cite{xie2022ng} to ensure 
its compatibility with LTE TDD. We pair a laptop with 
several USRP B210s \cite{USRPB210} to execute 
the \systemname{} decoder (\cref{fig:hw_impl:usrp_cell} \emph{(upper)}), 
where each USRP is decoding a distinct eNB to extract DCI. 
The extracted DCI provides input to our control 
program, coded in C. This program produces the requisite phase and filter 
configurations, which are then dispatched to the Arduino microcontrollers 
and DACs, to adjust the elements' phase shifter and filter.

\parahead{Experimental testbed}
We build a real Private LTE network in the CBRS band using 
a Sercomm 
\emph{Indoor Enterprise} CBRS Small Cell \cite{sercommCell} 
and the \emph{Aether} core
network software \cite{PetersonPrivate5G, aether}, shown 
in \cref{fig:hw_impl:usrp_cell} \emph{(lower)}. 
The network uses the Google SAS \cite{sas} to 
configure the operating channel of its eNBs.

%% file: text/7-eval.tex
\section{Evaluation}
\label{s:eval}

\begin{figure*}[htb]
    \begin{minipage}[b]{0.66\linewidth}
        \centering
        \includegraphics[width=0.99\textwidth]{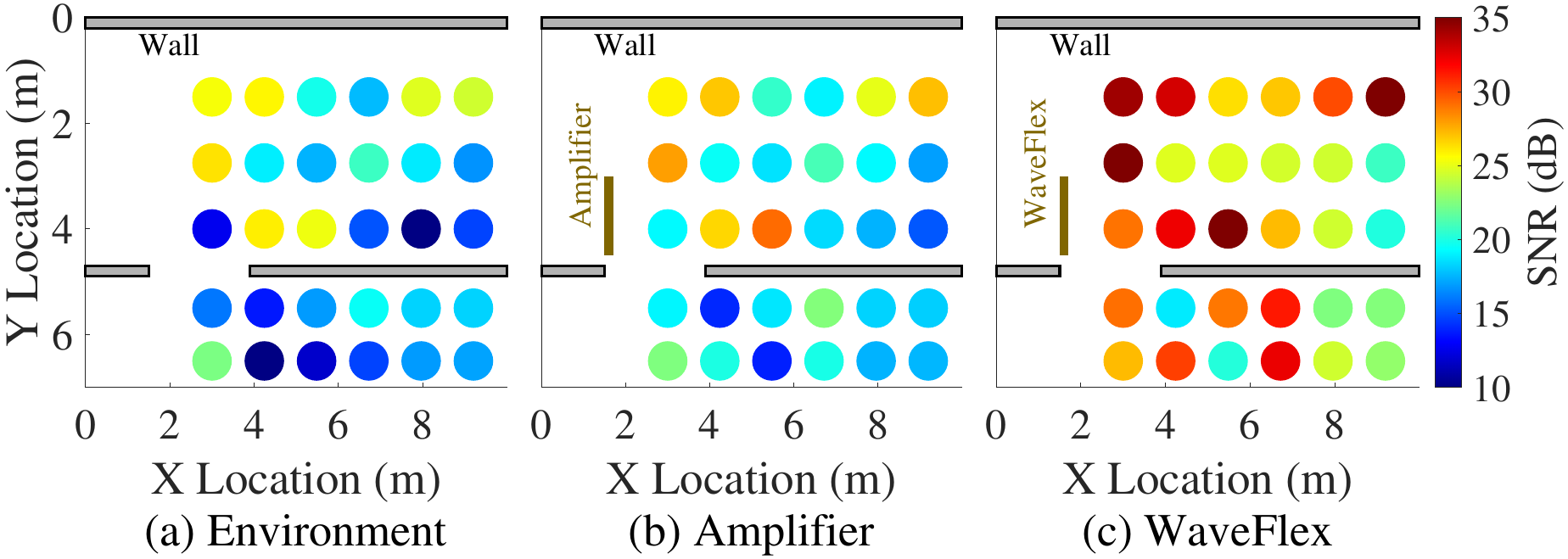}
        \caption{\textbf{SNR map} at 30 locations under three scenarios: (a) the environment alone, (b) the environment with amplifiers on, (c) \systemname{}.}
        \label{fig:eval_bench_30LocRoom}
    \end{minipage}    
    \hfill
    \begin{minipage}[b]{0.30\linewidth}
        \centering
        \includegraphics[width=0.95\textwidth]{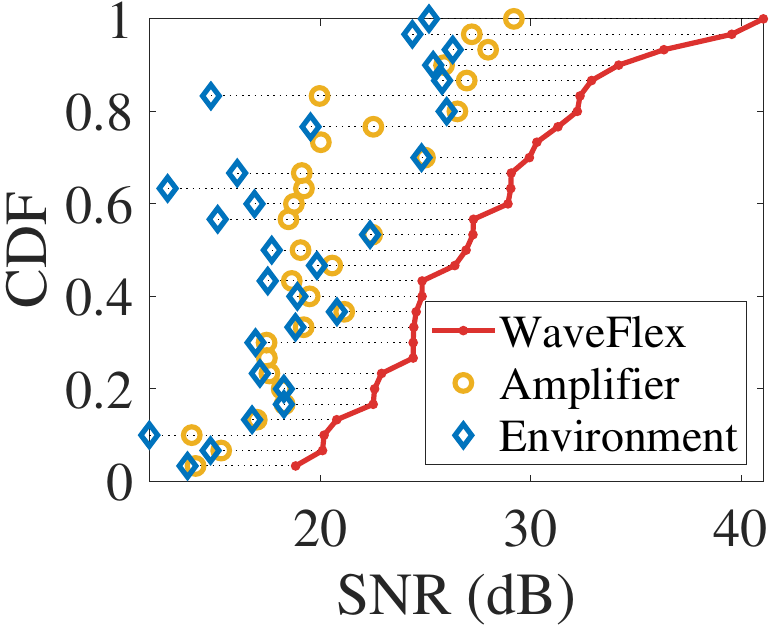}
        \caption{\textbf{Distribution of SNR} across 30 locations.}
        \label{fig:eval_bench_30LocCDF}
    \end{minipage} 
\end{figure*}

\begin{figure*}[htb]
    \begin{subfigure}{0.23\linewidth}
        \centering
        \includegraphics[width=0.99\textwidth]{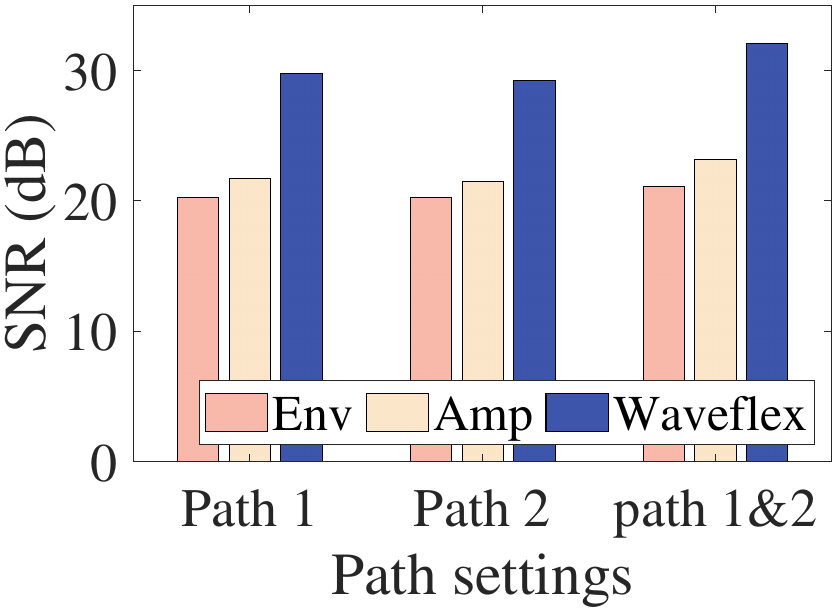}
        \caption{Across element paths.}
        \label{fig:eval_bench_2link}
    \end{subfigure}    
    \hfill
    \begin{subfigure}{0.40\linewidth}
        \centering
        \includegraphics[width=0.95\textwidth]{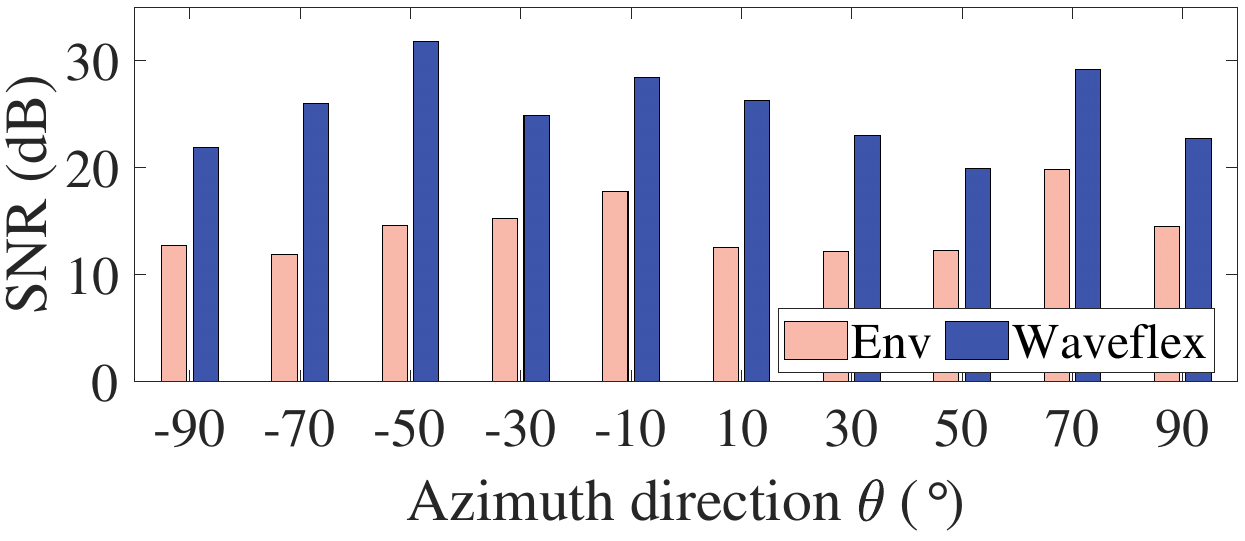}
        \caption{Across UE azimuthal bearing to \systemname{}.}
        \label{fig:eval_bench_azimuth}
    \end{subfigure} 
    \hfill
    \begin{subfigure}{0.36\linewidth}
        \centering
        \includegraphics[width=0.99\textwidth]{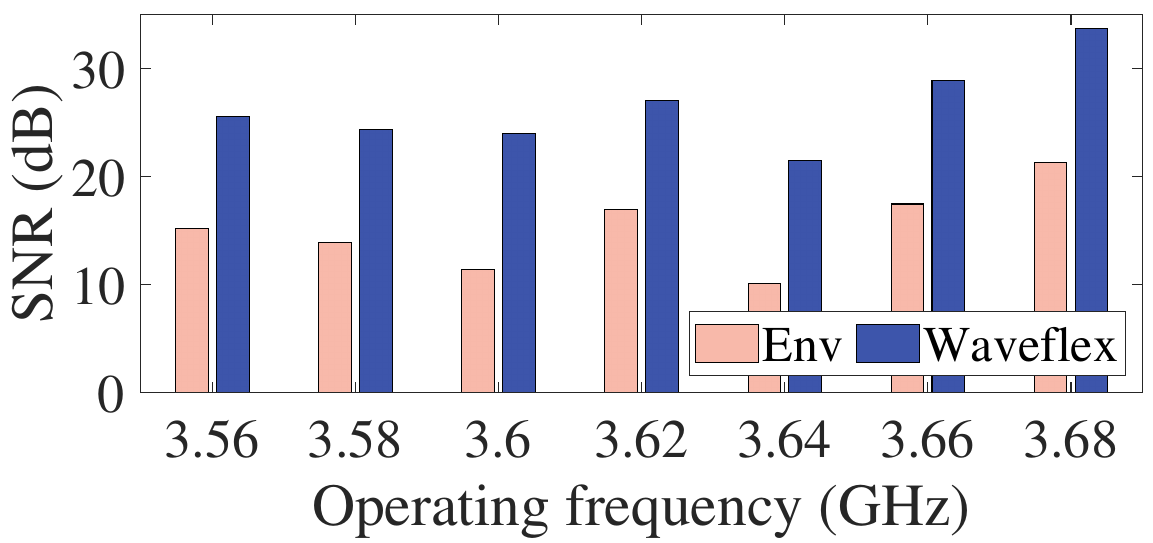}
        \caption{Across carrier frequency.}
        \label{fig:eval_bench_freq}
    \end{subfigure}     
    \caption{\textbf{Microbenchmark SNR measurements} stratified by element path, 
    azimuthal angle, and carrier frequency.}
\end{figure*}

In this section, we evaluate the performance of our \systemname{} implementation. 
We first 
introduce our evaluation methodology, and then present microbenchmarks 
evaluating the system under diverse conditions (\S\ref{s:eval:microbenchmarks}).
We then demonstrate \systemnames{} end\hyp{}to\hyp{}end performance in 
a real Private LTE network deployment, including a performance evaluation under 
dynamic conditions, including changes in mobility and traffic demand
(\S\ref{s:eval:e2e}).

\subsection{Experimental Methodology}
\label{s:eval:methodology}

\parahead{Microbenchmark experiments} 
For our filter measurement microbenchmarks, we use the Ansys HFSS
simulation software package and a Keysight E5063A Vector 
Network Analyzer (VNA) \cite{VNA}.
For our SNR measurement
microbenchmarks, we utilize two laptops equipped with 
USRP X310s, running SRS-ENB and SRS-UE, as the eNB and UE, respectively 
\cite{gomez2016srslte}. By default, the eNB operates at a 
frequency of 3.58~GHz.

Our evaluation metric for \systemname{}'s physical layer performance is the 
\emph{Signal-to-noise ratio (SNR)} at the UE side. 
We establish a baseline by measuring the SNR in the absence of 
\systemname{}. 
Additionally, we take a second baseline by powering\hyp{}on our 
surface amplifiers, but without executing the \systemname{} controller
to optimize performance. 
We evaluate \systemnames{} performance by comparing SNR when 
executing \systemname{}, to the SNR collected from these 
two baselines.

\parahead{End-to-end experiments}
We conduct end-to-end communication experiments on the testbed 
described in \cref{s:impl}. For the UE setup, we 
connect a Sercomm CBRS USB Dongle to a laptop~\cite{sercommUE}. 
Our experiments involve four eNBs and four UEs in total.
To decode DCI from four eNBs simultaneously, we connect four USRP B210s
to a laptop to run the \systemname{} decoder code.
We use the throughput measured at the UE side as the primary metric 
for these experiments.

\subsection{Microbenchmark Experiments}
\label{s:eval:microbenchmarks}

In this section, we evaluate \systemnames{} effectiveness in improving 
SNR under different conditions, to discern the source of 
improvements.

\subsubsection{Performance under single eNB}

\parahead{Performance across locations}
We fix the locations of the \systemname{} surface and eNB, and measure 
the SNR at 30 different locations to evaluate the SNR improvement of \systemname{}.
As shown in \cref{fig:eval_bench_30LocRoom}, we perform the 
experiment in an office scenario of area $10 \times 7$~$\mathrm{m}^2$.
The eNB is located at the left side of the surface, with a distance of five~m
from the surface.
We vary the locations of the UE (indicated by 
circles in \cref{fig:eval_bench_30LocRoom}) 
to cover diverse locations in our indoor office, 
varying distance and wall penetration.

At each location, the experiment is conducted 20 times. 
We calculate the average SNR over the 20 measurements and plot them in 
\cref{fig:eval_bench_30LocRoom}.
The figure shows results under three scenarios, without the \systemname{} 
surface, enabling amplifiers but without executing the \systemname{} 
controller, and enabling amplifiers and executing the \systemname{} controller.
From this figure, we can observe that in general \systemname{} is able to 
provide higher SNR improvements for close UEs. 
Compared with baseline, \systemname{} improves SNR by $9.12$~dB for 
the 15 closer locations, and improves SNR by $7.87$~dB for the 
15 farther locations on average.
For the ten through wall locations, \systemname{} can achieve an 
improvement of $8.60$~dB.
Comparing \cref{fig:eval_bench_30LocRoom}(b) and 
\cref{fig:eval_bench_30LocRoom}(c), we note that even though 
the amplifiers are able to achieve high SNR at some locations, 
\systemname{} outperforms the amplifier case at all locations. 
\cref{fig:eval_bench_30LocCDF} shows the CDF distribution 
of received SNR across 30 locations. 
On average, \systemname{} outperforms the amplifier case by $7.02$~dB.

\parahead{Performance across azimuthal bearing}
We measure \systemnames{} efficacy under 
different azimuthal bearings of the UE to the surface itself.
We separate the eNB and the \systemname{} surface by 5~m, and maintain 
the distance between the surface and the UE to 1~m. We vary UE 
locations to change the UE to surface azimuthal angle from $ -90^\circ$ to 
$90^\circ$ with a step size of $20^\circ$. 
\cref{fig:eval_bench_azimuth} presents the average SNR with 
different azimuth directions with and without the \systemname{} surface.
We can see that \systemname{} enhances SNR by $11.05$~dB on average 
across the entire azimuth range. 
\systemnames{} lowest improvement in this experiment is $8.26$~dB, which 
happens when at a $90^\circ$ bearing.

\parahead{Performance across element paths}
As introduced in \cref{s:design_hardware} on p.~\pageref{s:design_hardware},
our unit element comprises
two signal paths. We fix the locations of the eNB, the surface, and the UE, 
and measure the received SNR when only enabling Path~1, only 
enabling Path~2 and enabling both Path~1 and Path~2.
Results are shown in \cref{fig:eval_bench_2link}: 
compared to baseline, \systemname{} achieves a similar SNR improvement 
of $9.55$~dB and $8.94$~dB with only Path~1 and only Path~2, respectively.
Enabling both paths, \systemname{} is able to achieve a higher 
improvement of $10.94$~dB, 1--2~dB better than the respective paths individually.

\parahead{Performance across carrier frequency}
To validate \systemnames{} performance across the whole CBRS band, 
we conduct experiments that tune the carrier frequency of the
surface from 3.56~GHz to 3.68~GHz, with a step size of 20~MHz. 
We present the measured SNR with and without \systemname{}, 
in \cref{fig:eval_bench_freq}.
We observe that \systemname{} enhances the SNR for the 
entire CBRS band, achieving an average SNR improvement of $11.26$~dB.

\begin{figure}
    \centering
    \includegraphics[width=.7\linewidth]{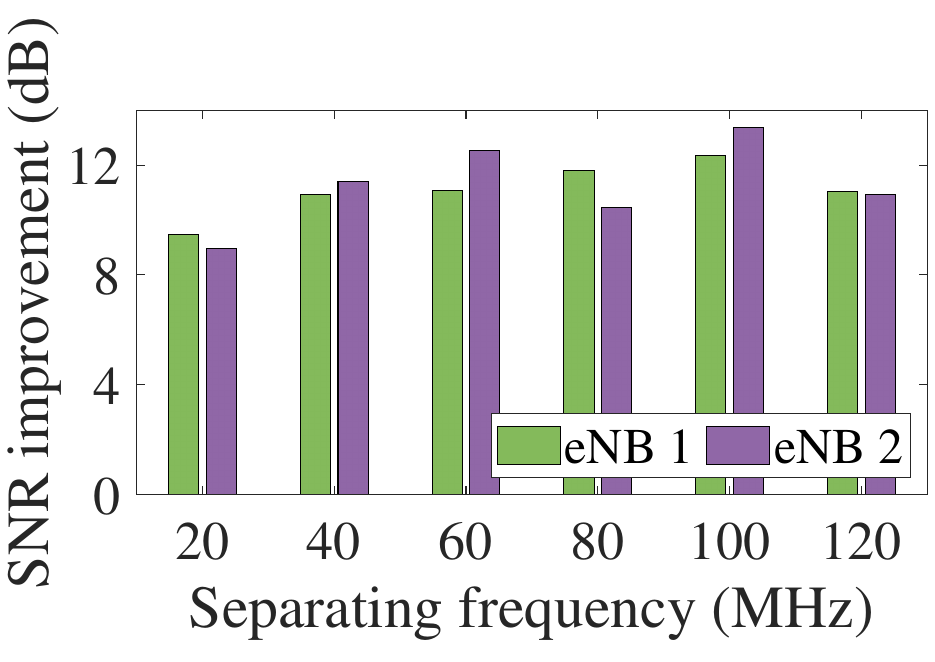}
    \caption{\systemnames{} SNR improvement of two eNBs versus
    different eNB carrier frequency separations.}
    \label{fig:eval_bench_freqGap}
\end{figure}

\begin{figure}[htb]
    \begin{subfigure}[b]{0.49\linewidth}
        \centering
        \includegraphics[width=\linewidth]{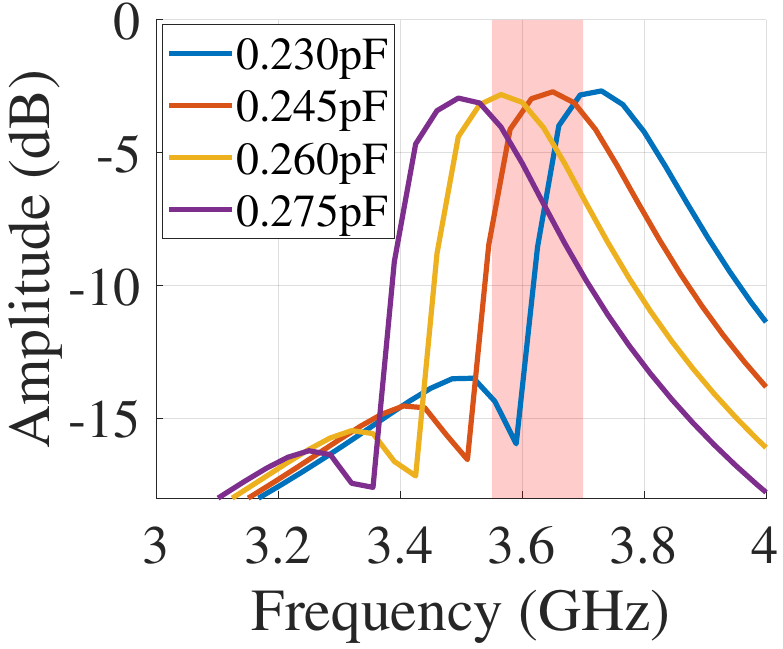}
        \caption{Amplitude response.}
        \label{fig:design_filter_simuAmp}
    \end{subfigure}
        \hfill
    \begin{subfigure}[b]{0.49\linewidth}
        \centering
        \includegraphics[width=\linewidth]{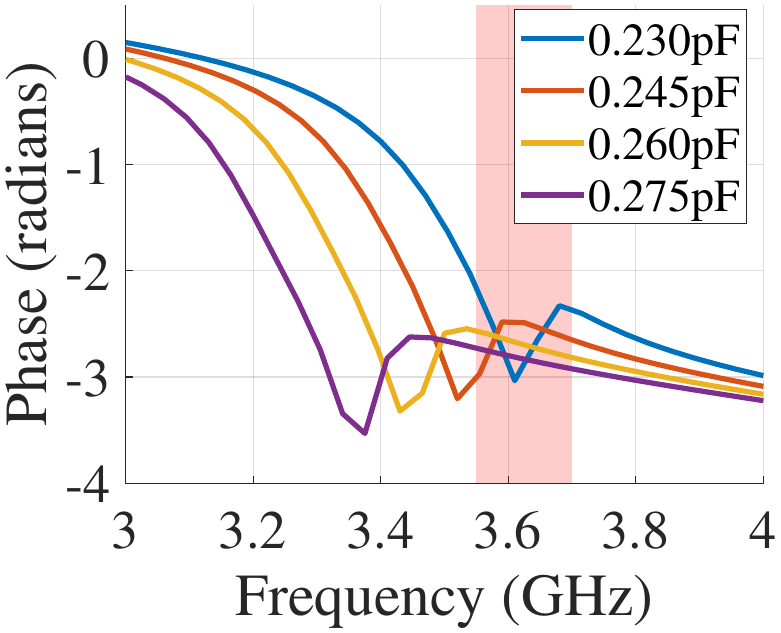}
        \caption{Phase response.}
        \label{fig:design_filter_simuPha}
    \end{subfigure}
    \caption{\textbf{Simulation: tunable filter} amplitude and phase response
    under varying varactor capacitances.}
    \label{fig:design_filter_simulation}
\end{figure}

\begin{figure}[htb]
    \begin{subfigure}[b]{0.49\linewidth}
        \centering
        \includegraphics[width=0.99\textwidth]{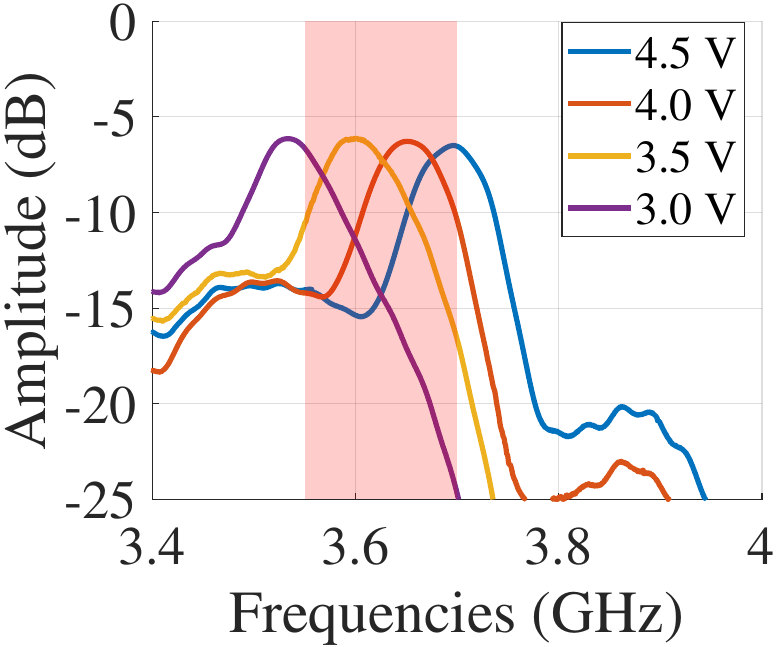}
        \caption{Amplitude response.}
        \label{fig:eval_bench_filterAmp}
    \end{subfigure}    
    \hfill
    \begin{subfigure}[b]{0.49\linewidth}
        \centering
        \includegraphics[width=0.99\textwidth]{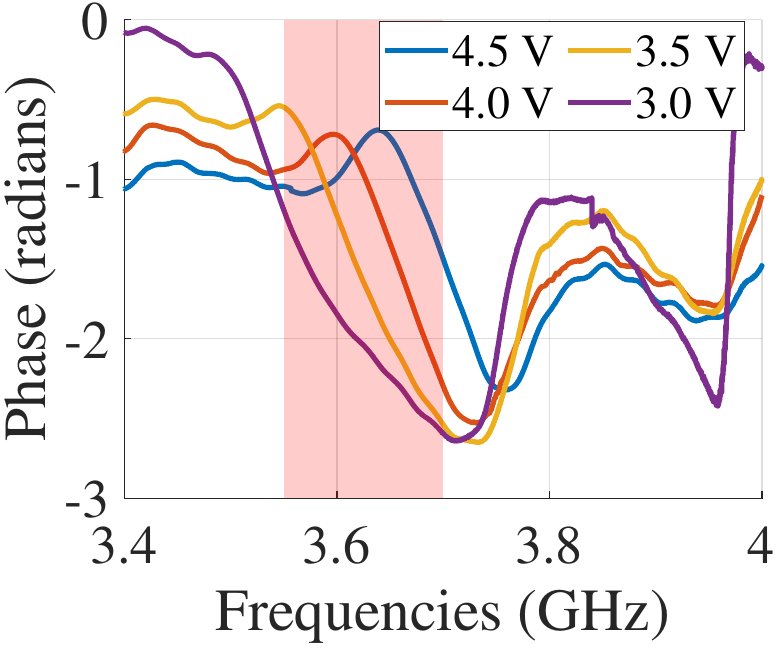}
        \caption{Phase response.}
        \label{fig:eval_bench_filterPha}
    \end{subfigure} 
    \caption{\textbf{VNA measurement: tunable filter} response versus frequency
    under different bias control voltages.}
    \label{fig:eval_vna_filter}         
\end{figure}

\subsubsection{Performance under Multiple eNBs}
We use two eNBs in this experiment to demonstrate \systemnames{} ability 
to simultaneously optimize multiple channels.

\parahead{Performance with different frequency separation}
To investigate how the frequency separation of two eNBs influences 
\systemnames{} multi-channel performance, we fix the frequency of eNB~1 at 
3.56~GHz, and change the frequency of eNB~2 from 3.58~GHz to 
3.68~GHz with a step size of 20~MHz.
The SNR improvement of eNBs~1 and~2 is presented in 
\cref{fig:eval_bench_freqGap}.
\systemname{} achieves SNR improvement of $9.49$~dB and $8.97$~dB for 
eNB 1 and eNB 2, respectively, when the separating frequency is 20~MHz.
At frequency separation exceeding 40~MHz, \systemname{} 
achieves an average SNR gain of $11.44$~dB and $11.74$~dB for eNB 1,
and eNB 2, respectively.

\subsubsection{Hardware verification}

This section presents simulated and then empirical (VNA\hyp{}based)
results on our filter performance.
We begin in simulation---\Cref{fig:design_filter_simulation} shows 
filter roll-off of up to 4~dB per 20~MHz
and an insertion loss between $-3.5$~dB and $-3.75$~dB 
when tuning the center frequency within the CBRS band. 
A phase response discontinuity occurs outside of the eNB's 
target frequency (20~MHz around the center frequency), and thus 
does not affect our phase tuning for the target channel.

To demonstrate \systemnames{} filter response and tunability across 
the CBRS band, we present our VNA measurements from 3.4 to 4~GHz in 
\cref{fig:eval_vna_filter}.
By tuning the bias voltage from 3.0 to 4.5~V, we are able to tune 
the center frequency from 3.55 to 3.7~GHz.
From \cref{fig:eval_bench_filterAmp}, we see that our fabricated 
filter exhibits an insertion loss of $-6.1$~dB, exceeding the simulated 
insertion loss (\emph{cf.}~\cref{fig:design_filter_simuAmp}) by 2.5~dB. 
We ascribe this increased loss to fabrication imperfections, encompassing 
factors such as transmission line loss and impedance mismatch.
The roll-off steepness is 3.23~dB per 20~MHz.

Turning now to empirical phase response, similar to the simulated phase response (\emph{cf.}~\cref{fig:design_filter_simuPha}),
we observe in \cref{fig:eval_bench_filterPha} that the phase discontinuity 
lies outside the eNB's target frequency range 
(20~MHz about the center frequency) thus 
it does not influence our phase tuning capability
within the target channel.

\begin{figure}[tb]
    \begin{subfigure}[b]{0.49\linewidth}
        \centering
        \includegraphics[width=0.99\textwidth]{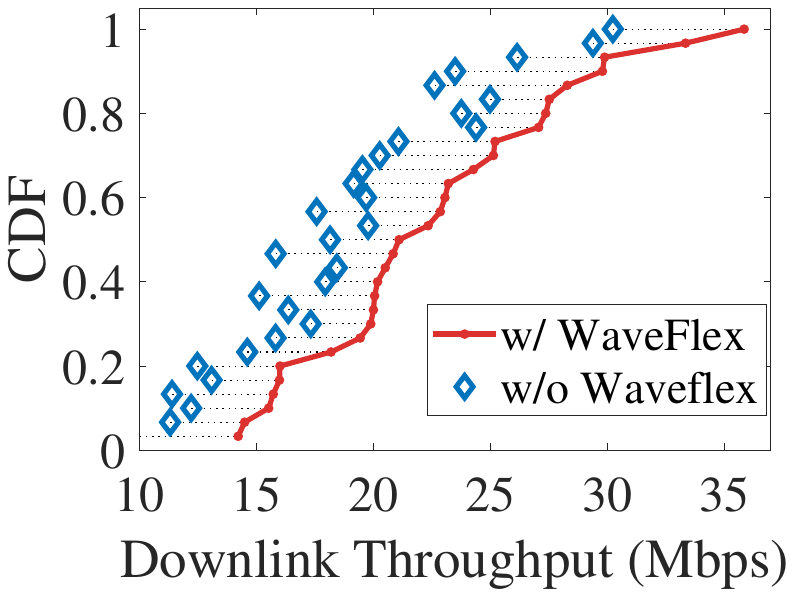}
        \caption{Downlink direction.}
        \label{fig:eval_e2e_30locDL}
    \end{subfigure}    
    \hfill
    \begin{subfigure}[b]{0.49\linewidth}
        \centering
        \includegraphics[width=0.99\textwidth]{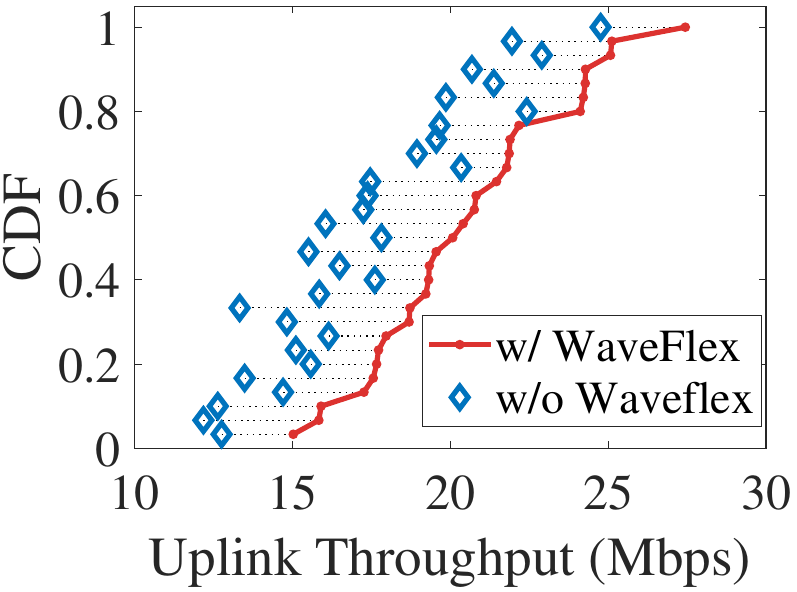}
        \caption{Uplink direction.}
        \label{fig:eval_e2e_30locUL}
    \end{subfigure} 
    \caption{\textbf{End-to-end performance (summary):} distribution 
    of downlink and uplink throughput with and without \systemname{} across 30 locations.}
    \label{fig:eval_e2e_30locCDF}         
\end{figure}

\begin{figure}[tb]
    \begin{subfigure}[b]{0.48\linewidth}
        \centering
        \includegraphics[width=0.99\textwidth]{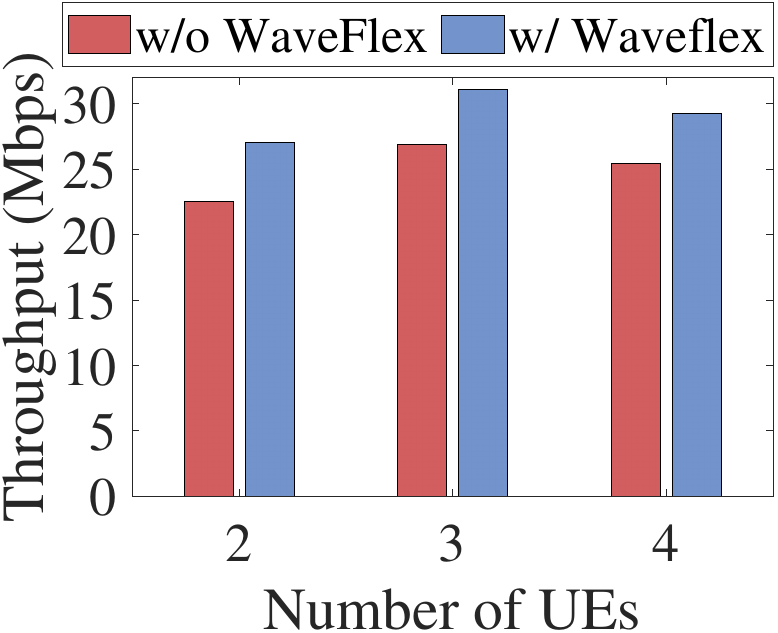}
        \caption{Varying \textbf{UE count.}}
        \label{fig:eval_e2e_nUE}
    \end{subfigure}    
    \hfill
    \begin{subfigure}[b]{0.48\linewidth}
        \centering
        \includegraphics[width=0.99\textwidth]{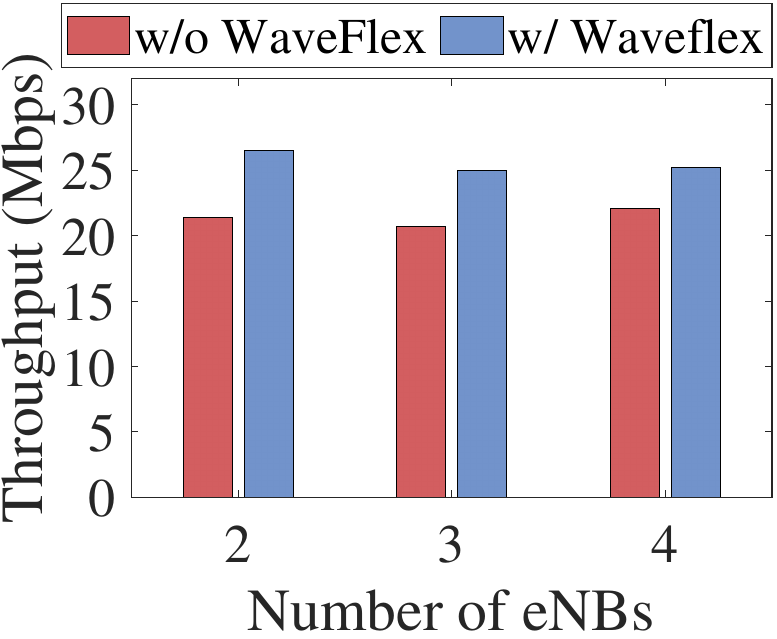}
        \caption{Varying \textbf{eNB count.}}
        \label{fig:eval_e2e_neNB}
    \end{subfigure} 
    \caption{\textbf{End-to-end performance (scaling):} 
    throughput per eNB, varying either UE or eNB count.}
    \label{fig:eval_e2e_container}
\end{figure}

\begin{figure*} 
    \centering
    \begin{subfigure}{0.35\textwidth} 
        \includegraphics[width=0.99\linewidth]{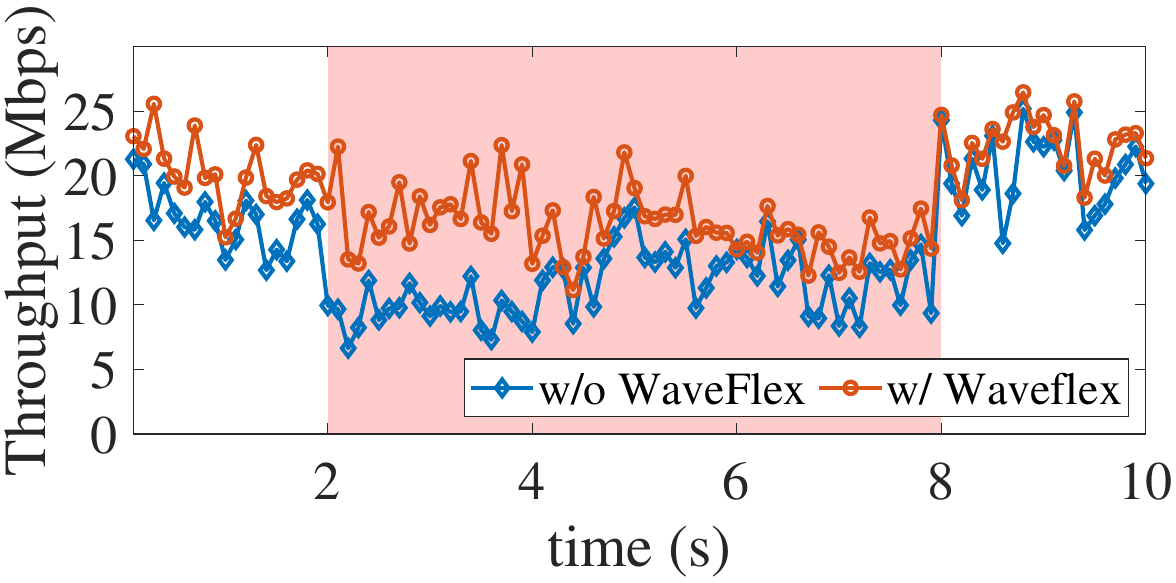}
        \caption{\textbf{Time series--} $v=0.5$~m/s.}
        \label{fig:eval_e2e_mobTrace}
    \end{subfigure}
    \hfill 
    \begin{subfigure}{0.21\textwidth}
        \includegraphics[width=0.99\linewidth]{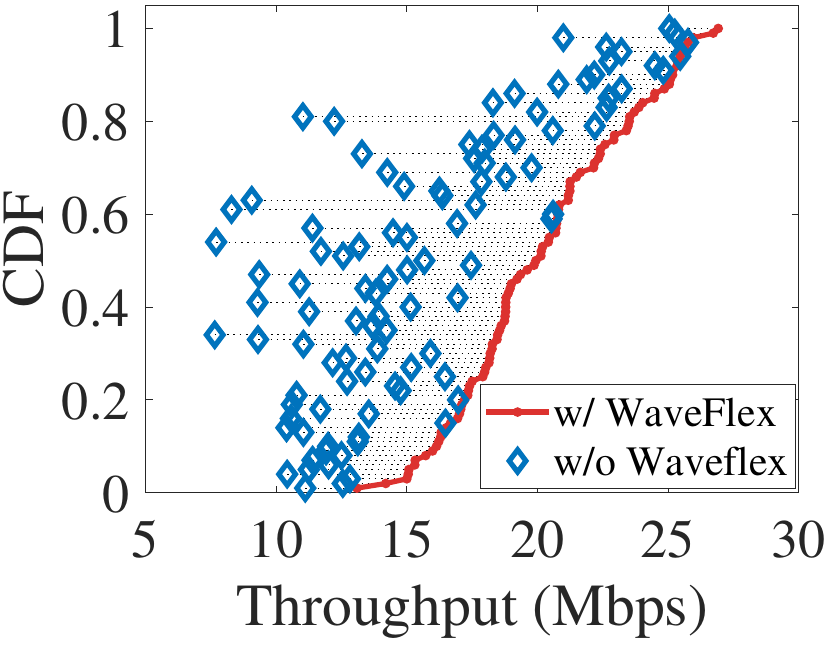}
        \caption{\textbf{CDF--} $v=0.2$~m/s. }
        \label{fig:eval_e2e_mobCDF_0R2m}
    \end{subfigure}
    \hfill 
    \begin{subfigure}{0.21\textwidth}
        \includegraphics[width=0.99\linewidth]{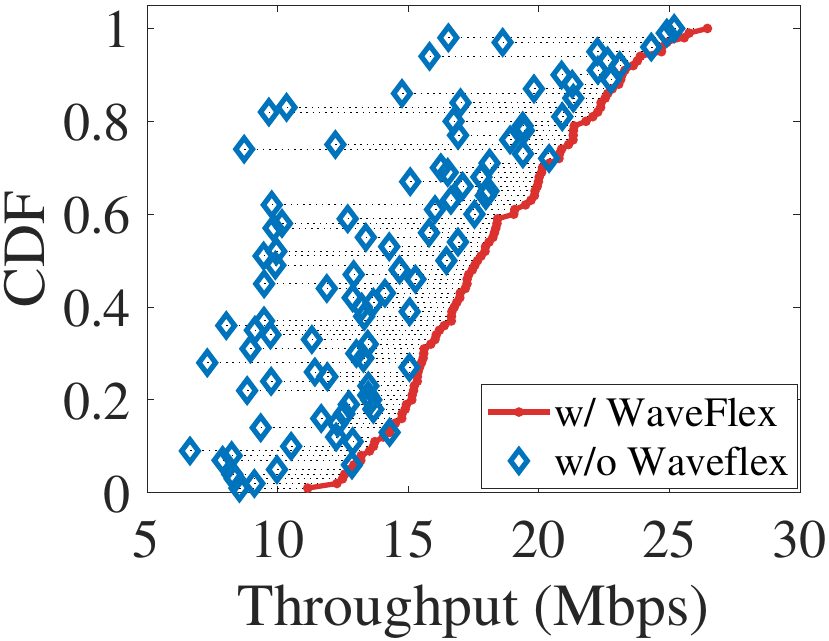}
        \caption{\textbf{CDF--} $v=0.5$~m/s.}
        \label{fig:eval_e2e_mobCDF_0R5m}
    \end{subfigure}
    \hfill 
    \begin{subfigure}{0.21\textwidth}
        \includegraphics[width=0.99\linewidth]{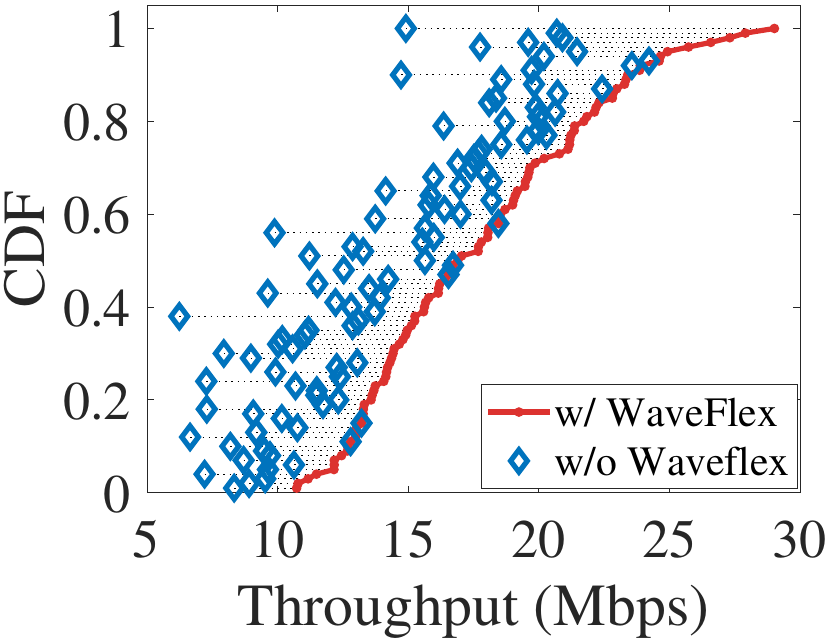}
        \caption{\textbf{CDF--} $v=1$~m/s.}
        \label{fig:eval_e2e_mobCDF_1m}
    \end{subfigure}
    \caption{\textbf{\systemnames{} performance under mobility:} (a) the throughput trace of 10 seconds with and without \systemname{} when UE moves with speed $v=0.5$~m/s; (b)-(d) the distribution of throughput with and without \systemname{} when UE moves with speed $v=0.2$~m/s, $v=0.5$~m/s, and $v=1$~m/s, respectively.}
    \label{fig:eval_e2e_mobility}
\end{figure*}


\begin{figure}[htb]
    \begin{minipage}[b]{0.95\linewidth}
        \centering
        \includegraphics[width=0.99\textwidth]{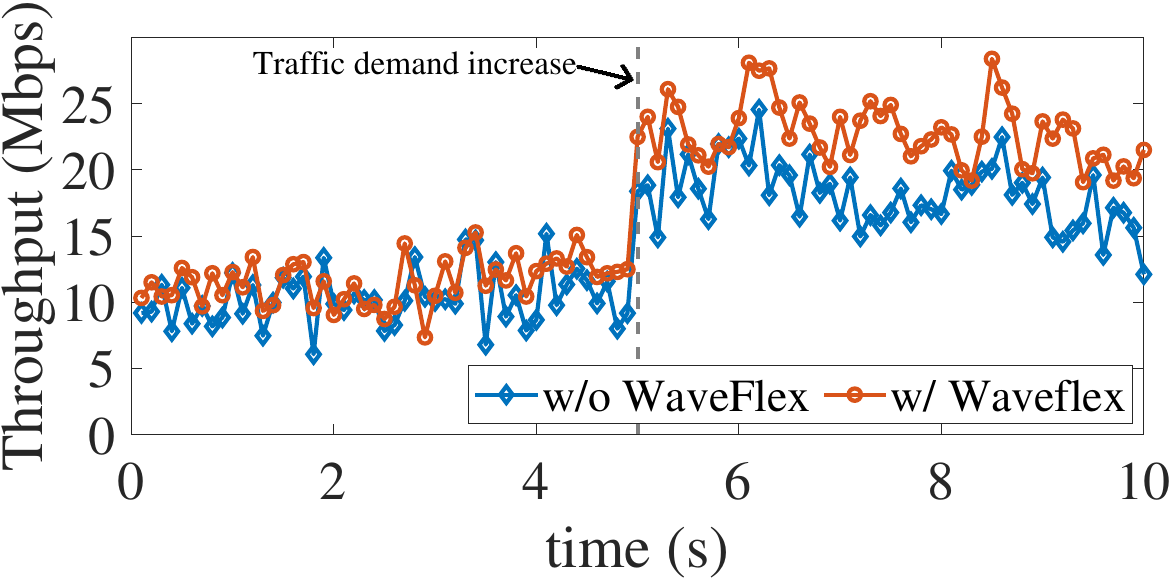}
        \caption{\textbf{\systemnames{} performance under traffic demand change:} the throughput trace of 10 seconds with and without \systemname{} when UE's traffic demand changes at the $5^{th}$ second.}
        \label{fig:eval_e2e_traffic}
    \end{minipage} 
\end{figure}

\begin{figure}[htb]
    \begin{subfigure}[b]{0.48\linewidth}
        \centering
        \includegraphics[width=0.99\textwidth]{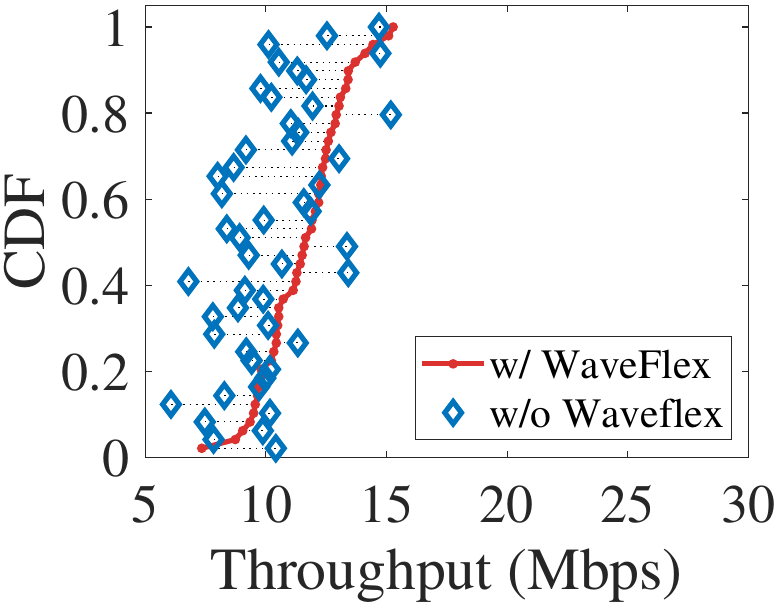}
        \caption{Before traffic change.}
        \label{fig:eval_e2e_trafficCDF_before}
    \end{subfigure}    
    \hfill
    \begin{subfigure}[b]{0.48\linewidth}
        \centering
        \includegraphics[width=0.99\textwidth]{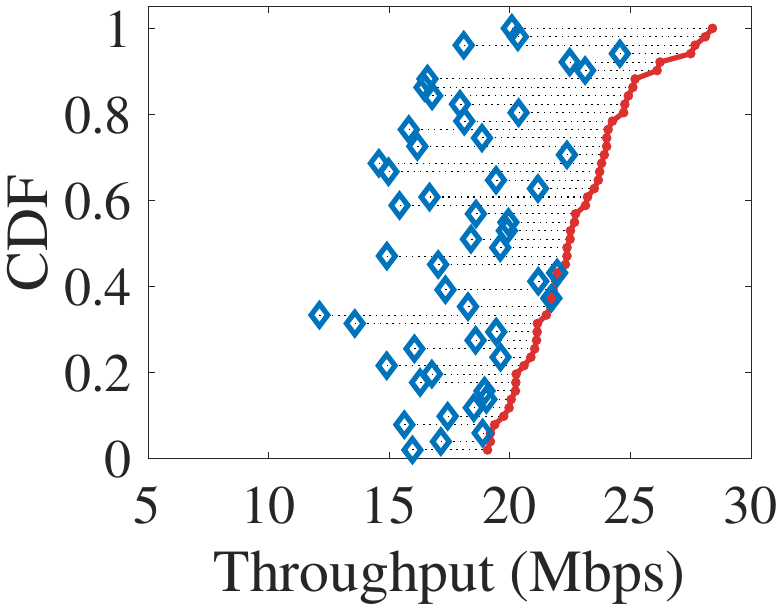}
        \caption{After traffic change.}
        \label{fig:eval_e2e_trafficCDF_after}
    \end{subfigure} 
    \caption{The distribution of throughput with and without \systemname{}  before and after the UE traffic demand changes.}
    \label{fig:eval_e2e_trafficCDF}
\end{figure}

\subsection{End-to-end Performance}
\label{s:eval:e2e}

In this section, we measure the end-to-end throughput under single eNB and 
multiple eNB scenarios. We also demonstrate the ability of \systemname{} to adapt 
in real time to wireless channel and traffic demand changes.

\subsubsection{Single eNB throughput}
\parahead{Single UE}
We fix the location of the \systemname{} surface, 5~m away from the 
eNB, and measure both downlink and uplink throughput at 30 different locations 
to evaluate \systemnames{} performance in end-to-end private LTE network.
We show the CDF distribution of downlink and uplink throughput across 30 
locations with and without \systemname{} in \cref{fig:eval_e2e_30locCDF}.
On average, \systemname{} is able to improve throughput by 4.36~Mbps in 
downlink, and improve throughput by 3.53~Mbps in uplink.

\parahead{Multiple UEs}
We change the number of UEs connected to the eNB from two to four, and measure 
the total downlink throughput of the eNB with and without \systemname{}, 
to demonstrate \systemnames{} efficacy:
\cref{fig:eval_e2e_nUE} shows that in a two\hyp{}UE scenario, 
\systemname{} improves the throughput by 4.50~Mbps on average. 
The throughput gain is 3.85~Mbps when there are four UEs under the 
same eNB.


\subsubsection{Multi-eNB throughput}
In this experiment, we investigate \systemnames{} capability of 
simultaneously optimizing for multiple eNBs.
We have four CBRS Small Cell eNBs located in two neighboring rooms, and we 
put the \systemname{} surface in between. 
We change the number of eNBs from two to four, and present 
the average throughput per eNB in \cref{fig:eval_e2e_neNB}. 
\systemname{} achieves a high per-eNB throughput gain of 5.10~Mbps with two 
eNBs, since the two paths design naturally fits in with the two eNB scenario.
When increasing the number of eNBs to three and four, 
\systemnames{} throughput gain drops to 4.32~Mbps and 3.19~Mbps, respectively.
We attribute such drops to the less number of elements allocated to each eNB, 
when the number of eNB increases.

\subsubsection{Mobility}
\label{s:eval_mobile}
To validate \systemnames{} performance in mobile scenarios, we show a 
trace of time-sequenced throughput when the UE moves along a specific 
route in \cref{fig:eval_e2e_mobility}. 
We fix the eNB to surface distance to 5~m, and initially 
the distance between the UE and the surface is 1~m. We then move the UE away 
from the surface at a constant speed of 0.5~m/s for 10~s. 
From 2--8~s (the highlighted part in \cref{fig:eval_e2e_mobility}), 
the UE to eNB line-of-sight (LoS) path is blocked.
We move the UE along this route twice to obtain throughput with and 
without \systemname{}.

During the whole 10~s process, \systemname{} achieves an average throughput 
gain of 3.95~Mbps, which shows that \systemname{} is able to keep up with 
environment changes.
From 2--8~s, \systemname{} improves the throughput by a higher 
value 4.58~Mbps, demonstrating \systemnames{} capability to provide higher 
throughput gain when the LoS path is blocked.

We present the distribution of throughput with and without \systemname{} 
when UE moves with different speeds in \cref{fig:eval_e2e_mobility}.
For UE speeds of $v=0.2, 0.5, 1$~m/s, the average throughput gains achieved 
by \systemname{} are 4.63~Mbps, 3.95~Mbps, and 3.3~Mbps, respectively. 
As the UE's speed increases, the \systemname{} controller detects channel 
changes more frequently, as depicted in \cref{s:design_BF_straw}. 
This results in fewer iterations for blind beamforming, subsequently 
leading to a reduced throughput gain.

\subsubsection{Traffic demand changes}
In this experiment, we use \emph{iperf} UDP to control the traffic demand of 
UE to test \systemnames{} performance under traffic demand changes.
We start by limiting the UE's traffic demand to 10~Mbps for 5~s, and then 
increase the traffic demand to 30~Mbps for another 5~s.

We show the time-sequenced throughput when UE's traffic demand changes with and 
without \systemname{} in \cref{fig:eval_e2e_traffic}. 
We also present the throughput distribution of \systemname{} before
and after a traffic demand change in \cref{fig:eval_e2e_trafficCDF}.
For the first 5~s, the throughput gain of \systemname{} is 1.38~Mbps, we argue 
that in this period, the wireless link capacity is enough to support the UE's
traffic even without \systemname{}, because the traffic demand is low.
From 5~s to 10~s, \systemname{} improves the throughput by 4.58~Mbps, which 
proves that \systemname{} is able to identify and react to traffic  demand 
changes.

%% file: text/9-concl.tex
\section{Conclusion}
\label{s:concl}

We have described the design and practical real-world implementation
of \systemname{}, the first smart surface that has demonstrated 
the capability to enhance 
the operation of real\hyp{}world 
Private LTE networks operating in the CBRS shared
spectrum licensing regime.  \systemname{} operates autonomously
from the core and RAN, easing deployment by obviating the need to
interface with cellular providers or even private cellular network
system administrators on-site.  The autonomous nature of 
our design opens up new possibilities to incrementally
deploy smart surfaces in private CBRS networks, and gives a new
direction to the area in general.  Also from an engineering
perspective, our design breaks new 
ground for smart surfaces,
adapting to shifting traffic demands and the vagaries of the 
wireless channel in real time.

\parahead{Limitations}  Our implementation and evaluation
are in a Private LTE network, and we acknowledge that 
implementation of \systemname{} for
Private 5G networks is ongoing and future work.
\systemname{} is designed for both Private LTE
and Private 5G CBRS networks, which have a slightly different 
control and data channel formats, but similar overall architecture. 
\systemname{} is implemented
in TDD\hyp{}mode LTE (as opposed to FDD mode), which the industry is
moving towards for 5G New Radio, owing to its superior performance.

%% file: text/10-acks.tex
\section{Acknowledgements}

This material is based upon work supported by the National Science Foundation under Grant Nos. CNS-2223556 and CNS-2232457 and the Sony Research Award Program.

%% file: text/appendix_file.tex
\section{Microstrip Filter Design}
\label{appen:microfilter}

We design a \emph{microstrip} filter, a low-cost but effective filter
that consists of a series of conductive metal traces printed on 
a PCB dielectric substrate.

\begin{figure}[t]
    \centering
    \begin{subfigure}[b]{0.29\linewidth}
        \centering
        \includegraphics[width=0.99\textwidth]{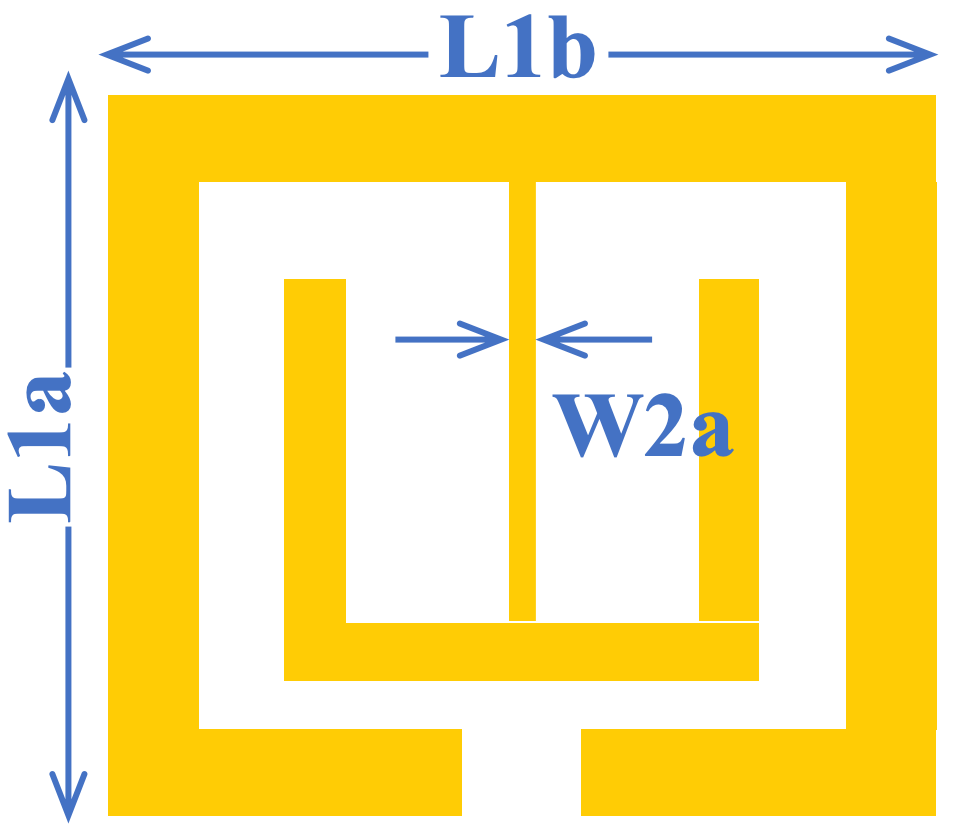}
        \caption{Design}
        \label{fig:design_openloop}
    \end{subfigure}
    \hfill    
    \begin{subfigure}[b]{0.69\linewidth}
        \centering
        \includegraphics[width=0.99\textwidth]{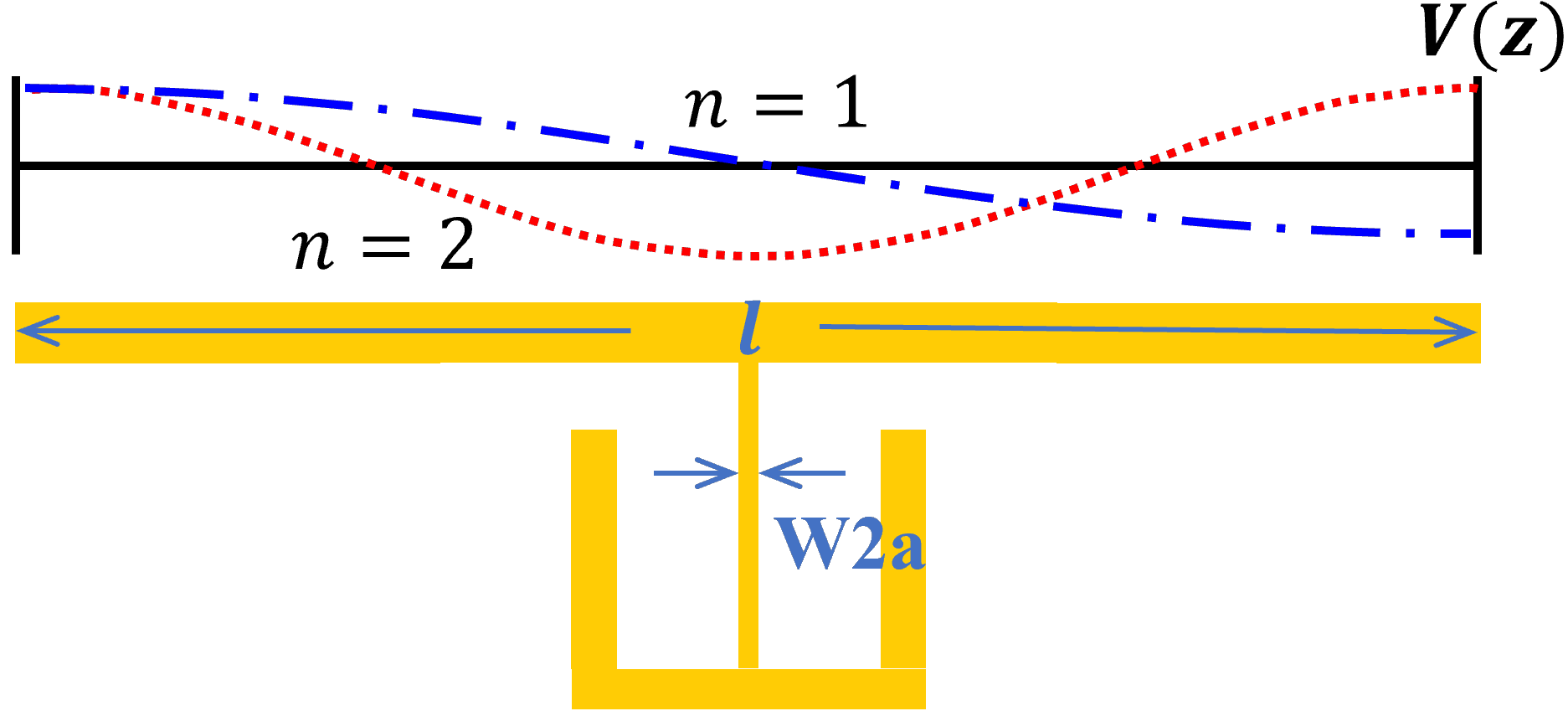}
        \caption{Voltage distribution.}
        \label{fig:design_V_distribute}
    \end{subfigure}
    \caption{\textbf{Filter design} and principle of operation.}
    \label{fig:design_filter_principle}
\end{figure}

The roll-off steepness of such filter depends on its order.\footnote{ The order of a filter depends on the number of reactive components (\textit{e.g.}, a third-order filter requires at least three components: one capacitor and two inductors, two capacitors and one inductor)}
A higher-order filter allows a steeper roll-off
but increases the design complexity and signal loss.
To mitigate this problem, \systemname{} uses a \emph{dual-mode open loop} 
microstrip filter as shown in \cref{fig:design_openloop}. Each of the two constituent resonator loops functions as a 
doubly-tuned resonant circuit, halving the number of 
required resonators for a higher-order filter \cite{hong2004microstrip} and thereby simplifying the filter design.

\Cref{fig:design_filter_principle} illustrates 
the operating principle of our filter. 
Consider unfolding the outer loop (\cref{fig:design_openloop}) into a straight line (\cref{fig:design_V_distribute}).
The loop resonates 
when its length $l = n \cdot \lambda / 2$ where $n$ is an arbitrary integer, and $\lambda$ is a wavelength.
In \Cref{fig:design_V_distribute}, we visualize the 
voltage distribution $V(z)$\footnote{The supply voltage is shared within the resonator loop. The sum of the voltages across the loop is equal to the voltage of the supply.} 
across the unfolded resonator when $n=1, 2$ (\textit{e.g.,} at its first two resonant frequencies). 
Since the resonator's two ends have a zero 
current, the voltage level maximizes at these two locations.
Hence, the curvatures of voltage distribution are restricted to two \textit{modes}.
%

The first mode is called \textit{an odd mode} with $n$ being an odd integer.
Here, the center of the outer resonator has a zero voltage.
Since the resonator cannot be excited at the location of voltage nulls,
the voltage does not pass into the inner loop, and the resonance occurs only in the outer loop.
The second mode is \textit{an even mode} with even $n$ and a maximum voltage level at the center.
Here, both the inner loading element and the outer loop resonate, 
giving rise to two separate resonant frequencies. 
By carefully adjusting the dimensions of the outer and inner loop,
we can tune the operating frequencies.
We design our resonator such that two resonant frequencies correspond to the edges of the desired pass band. 
We will further elaborate on our design parameters in \Cref{appen:filter}.

\section{Tunable Filter Dimensions}
\label{appen:filter}

We present a detailed derivation of tunable filter dimensions.
To concurrently and proportionally adjust the resonant frequencies of 
the odd and even modes by applying a single bias voltage, the tuning rates 
of both modes must be analyzed. The tuning rate reflects the change in 
a modal frequency resulting from a variation in the capacitance $C_v$. 
Assuming that when $C_v$ changes from $C_{v1}$ to $C_{v2}$, the odd-mode 
frequency shifts from $f^o_{01}$ to $f^o_{02}$, and the even-mode 
frequency alters from $f^e_{01}$ to $f^e_{02}$. The proportional tuning 
rate condition can be mathematically represented as,
\begin{equation}
R_t = \frac{f^o_{02}-f^o_{01}}{C_{v2}-C_{v1}} = \frac{f^e_{02}-f^e_{01}}{C_{v2}-C_{v1}},
\label{eqn:tune_rate}
\end{equation}
where $R_t$ denotes the tuning rate.

\begin{figure}[htb]
    \centering
    \includegraphics[width=0.7\linewidth]{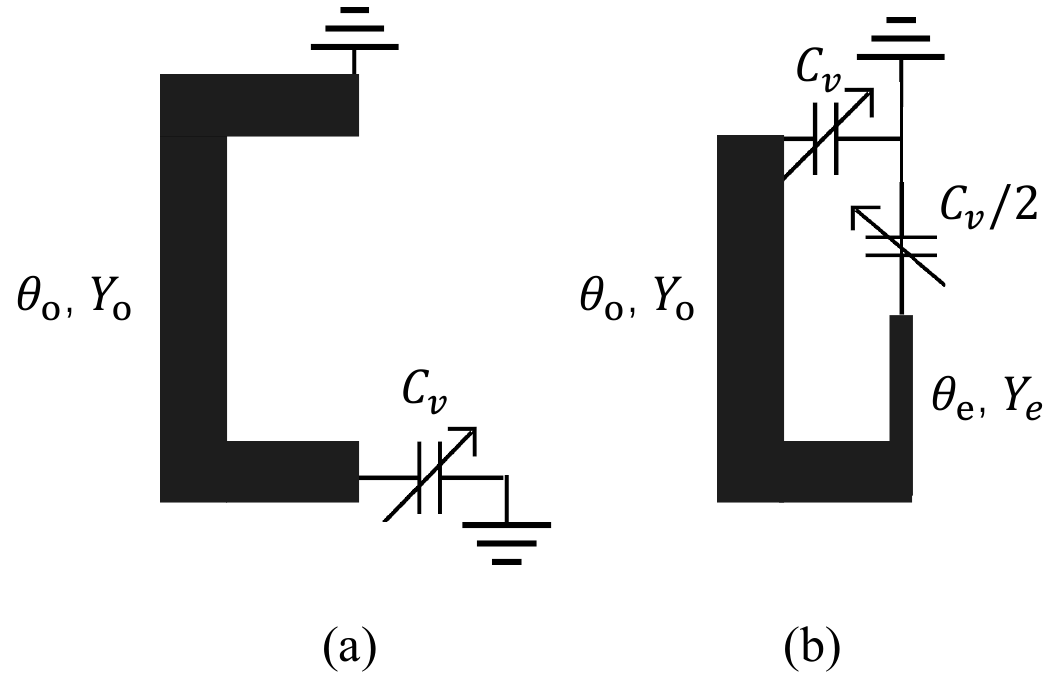}
    \caption{The circuit model of tunable filter: (a) odd mode, and (b) even mode.}
    \label{fig:appendix_model}
\end{figure}

The simplified circuit model for the odd and even modes is illustrated 
in Figure~\ref{fig:appendix_model}. $Y_o$, $Y_e$, $\theta_o$, 
$\theta_e$ represent the equivalent admittances and electrical lengths 
for the transmission line segments in the figure. 
According to \cite{hong2004microstrip}, the input admittance for the 
odd mode is given by,
\begin{equation}
Y_{ino} = j\left( \omega C_v-\frac{Y_o}{tan\theta_o}\right),
\label{eqn:admit_odd}
\end{equation}
and for the even mode, it is,
\begin{equation}
Y_{ine} = j\left( \omega C_v+Y_o\frac{Y_e+Y_otan\theta_o}{Y_o-Y_etan\theta_o}\right),
\label{eqn:admit_even}
\end{equation}
where $C_v$ represents the capacitance of the varactors. 
The resonant frequencies of the odd and even modes can be determined 
from the following conditions, respectively,
\begin{equation}
Im[Y_{ino}]=0, Im[Y_{ine}]=0.
\label{eqn:admit_resonant}
\end{equation}

By examining Equation~\ref{eqn:admit_odd} and 
Equation~\ref{eqn:admit_even}, it is apparent that the resonant 
frequency of the even mode can be altered by adjusting $Y_e$ and 
$\theta_e$, while keeping the resonant frequency of the odd mode 
unchanged. We solve the equations \ref{eqn:tune_rate}, 
\ref{eqn:admit_odd}, \ref{eqn:admit_even}, \ref{eqn:admit_resonant}
to obtain $Y_o$ and $\theta_o$, $Y_e$ and $\theta_e$, that align the 
tuning rate of the two modes.
$Y_{o/e}$ and $\theta_{o/e}$ correspond to the width and length of 
the metal traces, respectively.

\begin{figure}[htb]
    \centering
    \includegraphics[width=0.4\linewidth]{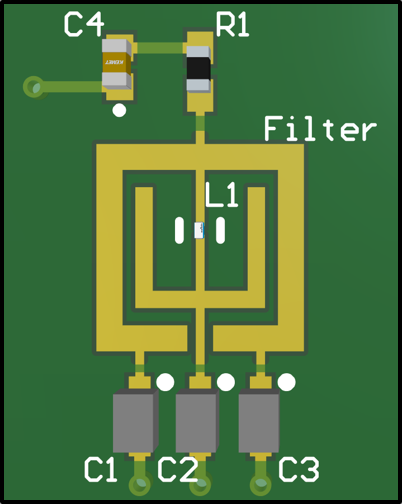}
    \caption{The 3D model of our tunable filter.}
    \label{fig:design_filter_3D}
\end{figure}

In order to ensure that the adjustable range encompasses the CBRS 
band from 3550~MHz to 3700~MHz, the resolved $Y_e$ value corresponds to 
a large inductance value, which equates to a thin loading element 
that cannot be fabricated. To overcome this issue, we incorporate 
an inductor L1~=~0.5~nH in the loading element, as shown in 
Figure~\ref{fig:design_filter_3D}.

%% file: main.bbl
\begin{thebibliography}{10}
\expandafter\ifx\csname urlstyle\endcsname\relax
  \providecommand{\doi}[1]{doi:\discretionary{}{}{}#1}\else
  \providecommand{\doi}{doi:\discretionary{}{}{}\begingroup
  \urlstyle{rm}\Url}\fi

\bibitem{aether}
{Aether: An ONF Project}.
\newblock \href{https://opennetworking.org/aether/}{[org]}.

\bibitem{dac}
{Analog Devices AD5370 40-ch. 16-bit DAC}.
\newblock
  \href{https://www.analog.com/media/en/technical-documentation/data-sheets/AD5370.pdf}{[pdf]}.

\bibitem{arduino}
{Arduino MKR Wi-Fi 1010}.
\newblock
  \href{https://docs.arduino.cc/resources/datasheets/ABX00023-datasheet.pdf}{[pdf]}.

\bibitem{arun2020rfocus}
V.~Arun, H.~Balakrishnan.
\newblock {RFocus: Beamforming using thousands of passive antennas}.
\newblock \textit{USENIX NSDI Symp.}, 1047--1061, 2020.

\bibitem{8796365}
E.~Basar, M.~Di~Renzo, J.~De~Rosny, M.~Debbah, M.-S. Alouini, R.~Zhang.
\newblock Wireless communications through reconfigurable intelligent surfaces.
\newblock \textit{IEEE Access}, \textbf{7}, 116,753--116,773, 2019.

\bibitem{9591503}
E.~Basar, H.~V. Poor.
\newblock Present and future of reconfigurable intelligent surface-empowered
  communications [perspectives].
\newblock \textit{IEEE Sig. Proc. Mag.}, \textbf{38}(6), 146--152, 2021.

\bibitem{cao2021reconfigurable}
X.~Cao, B.~Yang, H.~Zhang, C.~Huang, C.~Yuen, Z.~Han.
\newblock Reconfigurable intelligent surface-assisted {MAC} for wireless
  networks: protocol design, analysis, and optimization.
\newblock \textit{IEEE Internet of Things Journal}, 2021.

\bibitem{chen2020pushing}
L.~Chen, W.~Hu, K.~Jamieson, X.~Chen, D.~Fang, J.~Gummeson.
\newblock Pushing the physical limits of {IoT} devices with programmable
  metasurfaces.
\newblock \textit{USENIX NSDI Symp.}, 2021.

\bibitem{chen2023towards}
L.~Chen, B.~Yu, J.~Ren, J.~Gummeson, Y.~Zhang.
\newblock Towards seamless wireless link connection.
\newblock \textit{Proceedings of the 21st Annual International Conference on
  Mobile Systems, Applications and Services}, 137--149, 2023.

\bibitem{cho2022towards}
K.~W. Cho, Y.~Ghasempour, K.~Jamieson.
\newblock Towards dual-band reconfigurable metasurfaces for satellite
  networking.
\newblock \textit{Proceedings of the 21st ACM Workshop on Hot Topics in
  Networks}, 17--23, 2022.

\bibitem{dunna2020scattermimo}
M.~Dunna, C.~Zhang, D.~Sievenpiper, D.~Bharadia.
\newblock {ScatterMIMO: enabling virtual MIMO with smart surfaces}.
\newblock \textit{ACM MobiCom Conf.}, 2020.

\bibitem{USRPB210}
{The Ettus Research USRP B210}.
\newblock \href{https://www.ettus.com/all-products/ub210-kit/}{[url]}.

\bibitem{fcc35gHz}
FCC.
\newblock {3.5 GHz band overview}.
\newblock
  \href{https://www.fcc.gov/wireless/bureau-divisions/mobility-division/35-ghz-band/35-ghz-band-overview}{[fcc.gov]},
  2023.

\bibitem{gomez2016srslte}
I.~Gomez-Miguelez, A.~Garcia-Saavedra, P.~D. Sutton, P.~Serrano, C.~Cano, D.~J.
  Leith.
\newblock srs{LTE}: an open-source platform for {LTE} evolution and
  experimentation.
\newblock \textit{ACM WiNTECH}, 2016.

\bibitem{sas}
{Google Spectrum Access System (SAS)}.
\newblock \href{https://www.google.com/get/spectrumdatabase/sas/}{[url]}.

\bibitem{han2021dual}
J.~Han, R.~Chen.
\newblock Dual-band metasurface for broadband asymmetric transmission with high
  efficiency.
\newblock \textit{Journal of Applied Physics}, \textbf{130}(3), 2021.

\bibitem{hong2007dual}
J.-S. Hong, H.~Shaman, Y.-H. Chun.
\newblock Dual-mode microstrip open-loop resonators and filters.
\newblock \textit{IEEE Trans. on Microwave Theory and Techniques},
  \textbf{55}(8), 1764--1770, 2007.

\bibitem{hong2004microstrip}
J.-S.~G. Hong, M.~J. Lancaster.
\newblock \textit{Microstrip filters for RF/microwave applications}.
\newblock John Wiley and Sons, 2004.

\bibitem{jayaprakasam2017distributed}
S.~Jayaprakasam, S.~K.~A. Rahim, C.~Y. Leow.
\newblock Distributed and collaborative beamforming in wireless sensor
  networks: Classifications, trends, and research directions.
\newblock \textit{IEEE Comms. Sur. and Tut.}, \textbf{19}(4), 2092--2116, 2017.

\bibitem{VNA}
{E5063A ENA Vector Network Analyzer}.
\newblock
  \href{https://www.keysight.com/us/en/product/E5063A/e5063a-ena-vector-network-analyzer.html}{[url]}.

\bibitem{kumar2014lte}
S.~Kumar, E.~Hamed, D.~Katabi, L.~Erran~Li.
\newblock {LTE} radio analytics made easy and accessible.
\newblock \textit{ACM SIGCOMM CCR}, \textbf{44}(4), 211--222, 2014.

\bibitem{li2023rf}
X.~Li, C.~Feng, X.~Wang, Y.~Zhang, Y.~Xie, X.~Chen.
\newblock Rf-bouncer: A programmable dual-band metasurface for sub-6 wireless
  networks.
\newblock \textit{USENIX NSDI Symp.}, 2023.

\bibitem{li2019towards}
Z.~Li, Y.~Xie, L.~Shangguan, R.~I. Zelaya, J.~Gummeson, W.~Hu, K.~Jamieson.
\newblock Towards programming the radio environment with large arrays of
  inexpensive antennas.
\newblock \textit{USENIX NSDI Symp.}, 285--300, 2019.

\bibitem{visorsurf-commsmag2018}
C.~{Liaskos}, S.~{Nie}, A.~{Tsioliaridou}, A.~{Pitsillides}, S.~{Ioannidis},
  I.~{Akyildiz}.
\newblock {A New Wireless Communication Paradigm through Software-Controlled
  Metasurfaces}.
\newblock \textit{IEEE Comms. Mag.}, \textbf{56}(9), 162--169, 2018.

\bibitem{vmscatter-nsdi20}
X.~Liu, Z.~Chi, W.~Wang, Y.~Yao, T.~Zhu.
\newblock {VMscatter:} a versatile {MIMO} backscatter.
\newblock \textit{USENIX NSDI Symp.}, 895–910, 2020.

\bibitem{dirc-sigcomm09}
X.~Liu, A.~Sheth, M.~Kaminsky, K.~Papagiannaki, S.~Seshan, P.~Steenkiste.
\newblock {DIRC: Increasing Indoor Wireless Capacity Using Directional
  Antennas}.
\newblock \textit{Proc. of ACM SIGCOMM}, 2009.

\bibitem{9424177}
Y.~Liu, X.~Liu, X.~Mu, T.~Hou, J.~Xu, M.~Di~Renzo, N.~Al-Dhahir.
\newblock Reconfigurable intelligent surfaces: Principles and opportunities.
\newblock \textit{IEEE Comms. Surveys \& Tutorials}, \textbf{23}(3),
  1546--1577, 2021.

\bibitem{9377648}
R.~Long, Y.-C. Liang, Y.~Pei, E.~G. Larsson.
\newblock Active reconfigurable intelligent surface-aided wireless
  communications.
\newblock \textit{IEEE Trans. on Wireless Comms.}, \textbf{20}(8), 4962--4975,
  2021.

\bibitem{phaseShifter}
{Macom MAPS-010144 four-bit phase shifter}.
\newblock
  \href{https://www.mouser.com/datasheet/2/249/MAPS_010144-318345.pdf}{[pdf]}.

\bibitem{varactor}
{Macom MAVR-011005-12790T surface mount GaAs tuning varactor}.
\newblock
  \href{https://cdn.macom.com/datasheets/MAVR-011005-12790T.pdf}{[pdf]}.

\bibitem{amplifier}
{Mini-Circuits Low Noise Bypass Amplifier TSS-53LNB3+}.
\newblock \href{https://www.minicircuits.com/pdfs/TSS-53LNB3+.pdf}{[pdf]}.

\bibitem{splitter}
{Mini-Circuits Power Splitter/Combiner SCN-2-35+}.
\newblock
  \href{https://datasheet.octopart.com/SCN-2-35%2B-Mini-Circuits-datasheet-62339788.pdf}{[pdf]}.

\bibitem{denseAP}
R.~Murty, J.~Padhye, R.~Chandra, A.~Wolman, B.~Zill.
\newblock {Designing High Performance Enterprise Wi-Fi Networks}.
\newblock \textit{Proc. of USENIX NSDI Symp.}, 2008.

\bibitem{Peterson5G}
L.~Peterson, O.~Sunay.
\newblock \textit{{5G Mobile Networks: A Systems Approach}}.
\newblock {\href{https://github.com/SystemsApproach/5g}{[github]}}, 2022.
\newblock {License: CC BY-NC-ND 4.0}.

\bibitem{PetersonPrivate5G}
L.~Peterson, O.~Sunay, B.~Davie.
\newblock \textit{{Private 5G: A Systems Approach}}.
\newblock {\href{https://github.com/SystemsApproach/private5g}{[github]}},
  2022.
\newblock {License: CC BY-NC-ND 4.0}.

\bibitem{rotshild2021ultra}
D.~Rotshild, A.~Abramovich.
\newblock Ultra-wideband reconfigurable {X}-band and {Ku}-band metasurface
  beam-steerable reflector for satellite communications.
\newblock \textit{Electronics}, \textbf{10}(17), 2165, 2021.

\bibitem{saeidi202122}
H.~Saeidi, S.~Venkatesh, X.~Lu, K.~Sengupta.
\newblock 22.1 thz prism: One-shot simultaneous multi-node angular localization
  using spectrum-to-space mapping with 360-to-400ghz broadband transceiver and
  dual-port integrated leaky-wave antennas.
\newblock \textit{2021 IEEE International Solid-State Circuits Conference
  (ISSCC)}, vol.~64, 314--316. IEEE, 2021.

\bibitem{saifullah2021dual}
Y.~Saifullah, Q.~Chen, G.-M. Yang, A.~B. Waqas, F.~Xu.
\newblock Dual-band multi-bit programmable reflective metasurface unit cell:
  design and experiment.
\newblock \textit{Optics Express}, \textbf{29}(2), 2658--2668, 2021.

\bibitem{sercommCell}
{Sercomm Indoor Enterprise CBRS eNB}.
\newblock
  \href{https://www.sercomm.com/contpage.aspx?langid=1&type=prod3&L1id=2&L2id=1&L3id=107&Prodid=751}{[url]}.

\bibitem{sercommUE}
{Sercomm Indoor Enterprise CBRS UE}.
\newblock
  \href{https://www.sercomm.com/contpage.aspx?langid=1&type=prod3&L1id=2&L2id=2&L3id=110&Prodid=767}{[url]}.

\bibitem{tang2008compact}
W.~Tang, J.-S. Hong, Y.-H. Chun.
\newblock Compact tunable microstrip bandpass filters with asymmetrical
  frequency response.
\newblock \textit{Proc. of the European Microwave Conf.}, 599--602. IEEE, 2008.

\bibitem{tseng2011bio}
C.-S. Tseng, C.-C. Chen, C.~Lin.
\newblock A bio-inspired robust adaptive random search algorithm for
  distributed beamforming.
\newblock \textit{IEEE ICC Conf.}, 1--6, 2011.

\bibitem{tseng2014robust}
C.-S. Tseng, J.~Denis, C.~Lin.
\newblock On the robust design of adaptive distributed beamforming for wireless
  sensor/relay networks.
\newblock \textit{IEEE Trans. on Sig. Proc.}, \textbf{62}(13), 3429--3441,
  2014.

\bibitem{welkie2017programmable}
A.~Welkie, L.~Shangguan, J.~Gummeson, W.~Hu, K.~Jamieson.
\newblock Programmable radio environments for smart spaces.
\newblock \textit{ACM HotNets Workshop}, 2017.

\bibitem{xie2022ng}
Y.~Xie, K.~Jamieson.
\newblock {NG-Scope: Fine-grained} telemetry for {NextG} cellular networks.
\newblock \textit{Proc. of the ACM on Measurement and Analysis of Computing
  Systems}, \textbf{6}(1), 1--26, 2022.

\bibitem{yang-iet16}
B.~Yang, W.~Guo, Y.~Jin, S.~Wang.
\newblock Smartphone data usage: downlink and uplink asymmetry.
\newblock \textit{Electronics Letters}, \textbf{52}(3), 243--245, 2016.
\newblock \doi{https://doi.org/10.1049/el.2015.3249}.

\bibitem{9235486}
J.~Yuan, Y.-C. Liang, J.~Joung, G.~Feng, E.~G. Larsson.
\newblock Intelligent reflecting surface-assisted cognitive radio system.
\newblock \textit{IEEE Trans. on Comms.}, \textbf{69}(1), 675--687, 2021.

\bibitem{10.1145/3452296.3472890}
R.~I. Zelaya, W.~Sussman, J.~Gummeson, K.~Jamieson, W.~Hu.
\newblock {LAVA:} fine-grained {3D} indoor wireless coverage for small {IoT}
  devices.
\newblock \textit{ACM SIGCOMM Conf.}, 123–136, 2021.

\bibitem{10050148}
M.~Zeng, X.~Ning, W.~Wang, Q.~Wu, Z.~Fei.
\newblock {RIS aided NR-U and Wi-Fi coexistence in single cell and multiple
  cell networks on unlicensed bands}.
\newblock \textit{IEEE Trans. on Green Comms. and Net.}, \textbf{7}(3),
  1528--1541, 2023.

\bibitem{9998527}
Z.~Zhang, L.~Dai, X.~Chen, C.~Liu, F.~Yang, R.~Schober, H.~V. Poor.
\newblock Active {RIS} vs. passive {RIS}: Which will prevail in {6G}?
\newblock \textit{IEEE Trans. on Comms.}, \textbf{71}(3), 1707--1725, 2023.

\end{thebibliography}
